\documentclass[11pt]{article}
\usepackage{amsmath}
\usepackage{amssymb}
\usepackage{amsthm}
\usepackage{graphicx}
\usepackage[margin=1in]{geometry}
\usepackage{setspace}
\usepackage{caption}
\usepackage{booktabs}
\usepackage{multirow}
\usepackage{array}
\usepackage{hyperref}
\hypersetup{pdftex,colorlinks=true,allcolors=blue}
\usepackage{hypcap}

\providecommand{\keywords}[1]{\text{Keywords:} #1}

\title{Matching Algorithms for Causal Inference with Multiple Treatments}
\author{Anthony D. Scotina\footnote{Department of Mathematics and Statistics, Simmons University, Boston, MA 02115} \ and Roee Gutman\footnote{Department of Biostatistics, Brown University, Providence, RI 02912}}
\date{}

\begin{document}

\maketitle

\begin{abstract}
Randomized clinical trials (RCTs) are ideal for estimating causal effects, because the distributions of background covariates are similar in expectation across treatment groups. When estimating causal effects using observational data, matching is a commonly used method to replicate the covariate balance achieved in a RCT. Matching algorithms have a rich history dating back to the mid-1900s, but have been used mostly to estimate causal effects between two treatment groups. When there are more than two treatments, estimating causal effects requires additional assumptions and techniques. We propose several novel matching algorithms that address the drawbacks of the current methods, and we use simulations to compare current and new methods. All of the methods display improved covariate balance in the matched sets relative to the pre-matched cohorts. In addition, we provide advice to investigators on which matching algorithms are preferred for different covariate distributions. \\

\noindent
\end{abstract}

\section{Introduction}

\subsection{Overview}

Multi-arm randomized clinical trials have been proposed as an efficient trial design to compare multiple treatments simultaneously. The efficiency stems from the fact that comparisons between multiple treatments can be observed in one trial, rather than performing several trials each comparing only two treatments at a time \cite{barthel-09, parmar-14}. This efficiency is more pronounced when multiple treatments are compared to a control group, where multiple control groups would be required when conducting each two-arm trial separately. However, multi-arm trials can be more complex to design, conduct, and analyze \cite{ghosh-17}. One design complication is that in multi-arm trials all of the arms are required to follow a similar protocol with similar inclusion and exclusion criteria as well as primary and secondary outcomes. This is in contrast to multiple binary trials that may have different protocols for comparing each pair of treatments \cite{jaki-15}. Another design challenge with multi-arm trials is to ensure that there is no bias in the comparison of multiple treatments, which can be introduced through imbalances in units' allocations \cite{jaki-15}. A possible analysis complication is whether to make a statistical correction for the fact that multiple primary hypotheses are tested in the analysis \cite{wason-14}. Cook and Farewell (1996) \cite{cook-96} stated that when the examined hypotheses are prespecified, represent orthogonal contrasts, and are analyzed separately, then it is reasonable to not adjust for multiplicity. On the other hand, Bender and Lange (2001) \cite{bender-01} argue that experiments with multiple arms should always adjust for multiplicity when multiple significance tests are used as primary hypotheses in a confirmatory setting. 

These complications are exacerbated when comparing multiple treatments in non-randomized settings, because those receiving one treatment may differ from those receiving another with respect to the covariates, $\mathbf{X}$, which may also influence the outcome \cite{lopez-15}. The propensity score (PS), which is the probability of receiving treatment conditional on a set of observed covariates, is a common metric used in non-randomized observational comparisons to adjust for observed differences in covariates between two treatment groups. Matching on the propensity score has been shown in theory \cite{rosenbaum-83, rubin-92, rubin-06} and in application \cite{dehejia-02, monahan-10} to be a design technique that can generate sets of units with similar covariates' distributions on average. 

Propensity score matching is commonly employed using a greedy nearest-neighbor algorithm that does not guarantee optimal balance among covariates \cite{delosangelesresa-16}. Full matching was proposed a possible algorithm to  minimize the total sum of distances between matched units \cite{rosenbaum-89}. This algorithm is a special case of the minimum cost flow problem and can be solved in polynomial time. Cardinality matching is another matching method that maximizes the cardinality or size of the matched sample subject to constraints on covariates' balance \cite{zubizarreta-12}. Cardinality matching solves a linear integer programming problem which may be computationally intensive.

For more than two treatment groups, the generalized propensity score (GPS) vector represents each unit's probability of receiving any one of the treatments, conditional on the covariates \cite{imbens-00, imai-04}. Comparison of units with similar GPS vectors provides unbiased unit-level estimates of the causal effects between multiple treatments. While the literature on matching for estimating causal effects of binary treatments is extensive (\cite{zubizarreta-12, cochran-73, rubin-73a, rubin-73b, rosenbaum-02, hansen-04}, to name a few), generalizations and applications for multiple treatments remain limited, and currently no optimal matching algorithm has been proposed. Moreover, optimally matching for multiple treatments, also known as $k$-dimensional matching, was shown to be a NP-hard problem \cite{karp-72}, and the computational complexity of such methods may make them impractical in problems with many units and multiple treatments. 

In this paper, we propose several novel techniques for generating matched sets when estimating the causal effects between multiple treatments. We compare the performance of the new and previously proposed matching algorithms in balancing covariate distributions using simulation analyses. We also present a case study in which we compare the performance of different matching algorithms to estimate the effects of being released to different nursing homes from Rhode-Island hospital on patients propensity to be rehospitalized. 

\subsection{Framework}

For $Z$ possible treatment groups, $W_{i}\in\mathcal{W}=\{1,\dots,Z\}$ denotes the treatment group identification for unit $i\in\{1,\dots,n\}$. Let $n_{w}$ denote the size of treatment group $w$ such that $\sum_{w=1}^{Z}n_{w}=n$, and $T_{iw}$ is an indicator function that is equal to 1 if $W_{i}=w$ and to 0 otherwise. In addition, let $\mathcal{Y}_{i}=\{Y_{i}(1),\dots,Y_{i}(Z)\}$ be the set of potential outcomes for unit $i$, where $Y_{i}(w)$ is the potential outcome for unit $i$ if it were exposed to treatment $w$. In practice, only the potential outcome corresponding to the intervention that affected unit $i$ is observed. The other potential outcomes cannot be observed because they correspond to treatment assignments that did not occur \cite{rubin-74, rubin-78}. Assuming the Stable Unit Treatment Value Assumption (SUTVA) \cite{holland-86}, the observed outcome for unit $i$ can be written as
\[
	Y_{i}^{obs}=T_{i1}Y_{i}(1)+\cdots+T_{iZ}Y_{i}(Z).
\]

Because we cannot directly observe the causal effect for unit $i$, we need to rely on multiple units of which some are exposed to each of the other $Z-1$ possible treatments. For drawing causal inference, there are variables that are unaffected by $W_{i}$, the covariates $\mathbf{X}_{i}=(X_{i1}, \dots, X_{iP})$. A crucial piece of information that is needed for causal effect estimation is the assignment mechanism, which is the probability for each unit to receive one of the $Z$ treatments. If the $n$ units represent a random sample from an infinite super-population, then the assignment mechanism is individualistic \cite{lopez-15}, and it is unconfounded if 
\[
	P(W_{i}=w\mid \mathbf{X}_{i}, \mathcal{Y}_{i}, \phi)=P(W_{i}=w\mid \mathbf{X}_{i}, \phi)\equiv r(w,\mathbf{X}_{i}) 
\]
for all $i=1,\dots,n$ \cite{lopez-15}, where $r(w,\mathbf{X}_{i})$ is referred to as the generalized propensity score (GPS) for treatment $w$ and unit $i$ \cite{imai-04}, and $\phi$ is a vector parameter governing the distribution. The assignment mechanism is probabilistic if $0<P(W=w\mid \mathbf{X}, \mathcal{Y})$ for all $w$. We assume that the assignment mechanism is probabilistic and unconfounded throughout, so that comparing individuals with similar $R(\mathbf{X})\equiv (r(1,\mathbf{X}), \dots,r(Z,\mathbf{X}))$ results in well-defined causal effects.  

Commonly, $R(\mathbf{X}_{i})$ is unknown and only an estimate of it, $\hat{R}(\mathbf{X}_{i})$, is available. Thus, the observed data for unit $i$ comprise the vector $\{Y_{i}^{obs}, \mathbf{X}_{i}, W_{i}, \hat{R}(\mathbf{X}_{i})\}$. 

\subsection{Estimands}

Causal effect estimands are functions of unit-level potential outcomes on a common set of units \cite{rubin-74, rubin-78}. With $Z$ treatments, possible estimates of interest are the $\binom{Z}{2}$ pairwise population average treatment effects (ATE) between treatments $j$ and $k$, $\tau_{jk}\equiv E(Y(j)-Y(k))$, for $(j,k)\in\mathcal{W}^{2}$ and $j\neq k$, where expectation is taken over the entire population. The pairwise ATEs are transitive, such that $\tau_{ik}-\tau_{ij}=\tau_{jk}\ \forall\ i\neq j\neq k\in\mathcal{W}$. One possible extension of $\tau_{jk}$ would be to contrast treatments among a subset of units in the population receiving one of the $Z$ treatments and obtain the population average treatment effect on the treated (ATT) \cite{mccaffrey-13}, $\tau_{jk}^{t}\equiv E(Y(j)-Y(k)\mid W=t)$. For reference treatment group $W=t$, the ATTs are also transitive, such that $\tau_{ik}^{t}-\tau_{ij}^{t}=\tau_{jk}^{t}$, but transitivity of the ATTs does not extend to conditioning on different reference treatment groups. For example, $\tau_{ik}^{1}-\tau_{jk}^{2}$ is generally not equal to $\tau_{ij}^{1}$ \cite{lopez-15}. 

In observational studies, some units may have very low propensity to receive certain treatments. These units violate the positivity assumption of the assignment mechanism. One method to overcome this limitation is to restrict the analysis only to units that may receive all of the possible treatments. Lopez and Gutman (2017) \cite{lopez-15} proposed to include in the analysis only units that are in the rectangular common support region $\hat{r}(w,\mathbf{X})\in(\hat{r}_{min}(w,\mathbf{X}), \hat{r}_{max}(w,\mathbf{X}))\ \forall\ w\in\mathcal{W}$, where
\begin{align*}
	\hat{r}_{min}(w,\mathbf{X})&=\max\{\min(\hat{r}(w,\mathbf{X}\mid W=1)),\dots,\min(\hat{r}(w,\mathbf{X}\mid W=Z))\}\\
	\hat{r}_{max}(w,\mathbf{X})&=\min\{\max(\hat{r}(w,\mathbf{X}\mid W=1)),\dots,\max(\hat{r}(w,\mathbf{X}\mid W=Z))\}.
\end{align*}
 
Letting $E_{i}$ denote the indicator for all treatment eligibility:
\[
	E_{i}=
	\begin{cases}
	1,&\text{if}\ \hat{r}(w, \mathbf{X}_{i})\in(\hat{r}_{min}(w,\mathbf{X}_{i}), \hat{r}_{max}(w,\mathbf{X}_{i}))\ \text{for all}\ w\in\mathcal{W}\\
	0,&\text{otherwise}
	\end{cases}
\]
Then ATTs for units within this rectangular region are defined as 
\[
	\tau_{jk}^{t,E}=E(Y(j)-Y(k)\mid W=t, E_{i}=1).
\]
These estimands may restrict the analysis to a specific part of the population that may not be generalizable to the entire population who received treatment $t$. However, this restricted population is the one for whom all $Z$ treatment have positive probability of being administered and for which an equipoise between all $Z$ treatment exists.  In that sense, these estimands sacrifice some external validity to improve the internal validity \cite{imbens-15}. In many causal inference questions, having a credible and precise answer for a subpopulation is considered more important than having a controversial imprecise answer for the entire target population \cite{imbens-15}.

The estimands $\tau_{jk}$ and $\tau_{jk}^{t}$ can be approximated using the sample average treatment effects:
\[
	\hat{\tau}_{jk}\equiv\frac{1}{n}\sum_{i=1}^{n}(Y_{i}(j)-Y_{i}(k)),\quad\hat{\tau}_{jk}^{t}\equiv\frac{1}{n_{t}}\sum_{i=1}^{n}T_{it}(Y_{i}(j)-Y_{i}(k)). 
\]
These approximations are hypothetical, because one can only observe one of the potential outcomes for each unit. Matching procedures have been proposed as one possible solution to this problem \cite{lopez-15,yang-16}.

\subsection{Distance measures} \label{subsec::distance.measures}

To define the similarity between patients receiving different treatments, a distance measure is required. The linear GPS is one distance measure that can be used in the multiple treatment setting, 
\[
	|\text{logit}(r(w,\mathbf{X}_{i}))-\text{logit}(r(w,\mathbf{X}_{j}))|,\quad i\neq j\in\{1,\dots,n\},\ w\in\{1,\dots,Z\}.
\]
Using the transformed GPS is preferred to a distance measure with the untransformed $r(w,\mathbf{X}_{i})$ because matching on the logit transformation has been shown to produce lower bias in matched samples \cite{lopez-15, rosenbaum-85, rubin-01}. One limitation of this distance is that it only compares one component of the GPS at a time. Matching only on a single component may not necessarily result in similar values for all of the other components of the GPS vector. This limitation is exacerbated as the number of treatment groups increases.

The Euclidean distance is a multivariate matching metric that is the sum of the squared differences between vector's components. Formally, the Euclidean distance between vectors $\mathbf{V}_{i}=(v_{i1},\dots,v_{iP})$ and $\mathbf{V}_{j}=(v_{j1},\dots,v_{jP})$ is 
\[
	\{(\mathbf{V}_{i}-\mathbf{V}_{j})^{\text{T}}(\mathbf{V}_{i}-\mathbf{V}_{j})\}^{1/2}=\sqrt{\sum_{k=1}^{P}(v_{ik}-v_{jk})^{2}}, 
\]
where $P$ is the dimension of $\mathbf{V}_{i}$. A limitation of the Euclidean distance as a matching metric is its sensitivity to the correlation structure of the vector's components \cite{krzanowski-00, steiner-13}. 

This sensitivity to the correlation of covariates is mitigated with the Mahalanobis distance \cite{cochran-73, rubin-79, rubin-80}. The Mahalanobis distance between $\mathbf{V}_{i}$ and $\mathbf{V}_{j}$ is
\[
	\{(\mathbf{V}_{i}-\mathbf{V}_{j})^{\text{T}}\boldsymbol\Sigma^{-1}(\mathbf{V}_{i}-\mathbf{V}_{j})\}^{1/2}, 
\]
where $\boldsymbol\Sigma$ is the covariance matrix of $\mathbf{V}_{i}$. Commonly, $\boldsymbol\Sigma$ is unknown and an estimate of it, $\hat{\boldsymbol\Sigma}$, is used instead. With binary treatment, matching on the Mahalanobis distance of the covariates performs well in reducing covariate bias between the two treatment groups when the covariates' space is small. However, the reduction in bias is less than optimal when the covariates are not normally distributed or there are many covariates \cite{gu-93, stuart-10}. When defining similarity of units with multiple treatments, the Euclidean and the Mahalanobis distances can be applied to either the original covariates, the untransformed estimated GPS vector, or the transformed estimated GPS vector. Because the Mahalanobis distance is frequently preferred to the Euclidean distance, the latter distance was not examined in our simulation analysis. 

\section{Matching algorithms}\label{sec:matching}

\subsection{Basic matching algorithms} \label{subsec:simple.matching}

Matching procedures for causal inference attempt to find units that are ``close" to each other in terms of $\mathbf{X}$, but receive a different treatment. It is relatively straightforward to perform matching when only a single covariate influences the assignment to treatment, but this task can become more complex as the number of covariates increases. 

The basic matching with replacement algorithm identifies for unit $i$, the units with the shortest distances from each of the other treatment groups. Because this algorithm identifies matches to all of the units, some of the matches may not be very close in terms of the distance measure. One possible solution is to restrict all of the matches to have a distance that is smaller than a pre-defined threshold (caliper). 

The basic matching algorithm with reference treatment $t\in\mathcal{W}$ is summarized as follows: 
\begin{enumerate}
	\item Estimate $R(\mathbf{X}_{i})$, $i=1,\dots,n$ using a multinomial logistic regression model. 
	\item Drop units outside the common support (e.g., those with $E_{i}=0$), and re-fit the model once. 
	\item For all $t'\neq t$, match those receiving $t$ to those receiving $t'$ using a pre-specified distance measure and a caliper. 
	\item Units receiving $t$ who were matched to units receiving all treatments $w\neq t$, along with their matches receiving the other treatments, comprise the final matched cohort. 
\end{enumerate}

Yang et al. \cite{yang-16} have implemented this procedure with the Mahalanobis distance of the covariates and the linear GPS in step 4, but with a slight modification to step 2. Instead of removing all units with $E_i =0$, they proposed to keep only units for which $\sum_{w=1}^{Z} \frac{1}{r(w,\mathbf{X}_i)} \leq \lambda$, where $\lambda$ is a predefined threshold parameter. Under the assumptions of constant treatment effects and homoscedasticity, threshold $\lambda$ is estimated such that the variances of each pairwise treatment effect are minimized. A possible limitation of the trimming procedures is that units with different $R(\mathbf{X}_{i})$ may be dropped, resulting in dropped units with different covariates’ distributions. Thus, we compared all of the methods only to units with $E_i=1$.

\subsection{Vector matching (VM)} \label{subsec:vm}

The basic matching algorithm relies on a distance measure that aggregates individual component differences over the entire vector. In some cases, it may result in some components of the matched vector that are far apart and other components that are relatively close. Vector matching (VM) is a possible algorithm for matching units in observational studies with multiple treatments that addresses this limitation \cite{lopez-15}. VM uses $k$-means clustering to group units such that within each cluster, units are roughly similar on $Z-2$ components of the GPS, and it performs the matching on the remaining component using the linear GPS only among units that are in the same cluster. The matching mehthod proposed by Yang et al. \cite{yang-16} is a special case of VM when $k=1$. However, for $k>1$, VM provides a valid estimate of the average treatment effects because it ensures that units which are matched to one another are nearly perfect on one component of the GPS and roughly similar on the other components \cite{lopez-15}.  

For a reference treatment $t\in\mathcal{W}$, the VM procedure is summarized as follows:
\begin{enumerate}
	\item Estimate $R(\mathbf{X}_{i})$, $i=1,\dots,n$ using a multinomial logistic regression model. 
	\item Drop units outside the common support (e.g., those with $E_{i}=0$), and re-fit the model once. 
	\item For all $t'\neq t$:
	\begin{enumerate}
		\item Partition all units using $k$-means clustering on $\text{logit}(\hat{R}_{t,t'}(\mathbf{X}))\equiv(\text{logit}(\hat{r}(w,\mathbf{X}))\ \forall\ w\neq t,t')$. This forms $K$ strata of units with relatively similar components of the GPS vectors. 
		\item Within each strata $k\in \{1,\dots,K\}$, match all units receiving $t$ to those receiving $t'$ on $\text{logit}(\hat{r}(t,\mathbf{X}_{i}))$ with replacement using a caliper of $\epsilon*\text{sd}(\text{logit}(\hat{r}(t,\mathbf{X}_{i})))$, where $\epsilon=0.5$. 
	\end{enumerate}
	\item Units receiving $t$ who were matched to units receiving all treatments $w\neq t$, along with their matches, comprise the final matched cohort. 
\end{enumerate}

VM was shown to yield the lowest average maximum pairwise bias in comparison to other nearest neighbor matching and weighting methods for multiple treatments, while retaining a high percentage of units matched \cite{lopez-15}.

VM has several potential limitations. First, it is a 1:1:$\cdots$:1 nearest neighbor matching algorithm. While this ensures that each unit will be paired with its best match, this could leave units outside the final matched cohort, resulting in larger sampling errors. However, the increase in sampling error is usually minimal as long as the reference group stays the same size and only the compared groups decrease in size \cite{cohen-88, ho-07}. In addition, the gain in power may be hindered by increasing bias that arises from the introduction of poorer matches. Second, using an identical caliper for all units could result in significant loss of units in the reference group that are left without matches in the other $Z-1$ groups. Third, when the number of treatments increases, it is harder for VM to ensure that the matched sets are balanced across all components of the GPS vector, because it only performs close matching on one component of $R(\mathbf{X})$. Fourth, because VM uses $k$-means clustering to classify units in step 3(a), possible matches that are on the boundaries of different clusters may not be considered, which may result in increased bias.

\subsection{Extensions to VM} \label{subsec:vm.extensions}

We examine several variations of VM that attempt to address the limitations described in Section~\ref{subsec:vm}. To address the first limitation, we examined VM with two matches for each unit (e.g., 1:2:2 matching for $Z=3$) (VM2), and VM using matching without replacement in step 3(b) (VMnr). To address the second limitation, we examined omitting the caliper in step 3(b) (VMnc). To address the third limitation of VM, we examined using the Mahalanobis distance of $\left\{\text{logit}(\hat{r}(t,\mathbf{X})),\text{logit}(\hat{r}(t',\mathbf{X}))\right\}$ in step 3(b) with and without a caliper (KM and KMnc). To address the fourth limitation of VM, we used fuzzy clustering instead of $k$-means clustering in step 3(a) (VMF). To address limitations 2-4 simultaneously, we combined fuzzy clustering in step 3(a) with the Mahalanobis distance of $\left\{\text{logit}(\hat{r}(t,\mathbf{X})),\text{logit}(\hat{r}(t',\mathbf{X}))\right\}$in Step 3(b) with and without a caliper (FM and FMnc).

In contrast to $k$-means clustering which assigns units to only one cluster, fuzzy clustering assigns each unit a probability to be in a cluster. A common fuzzy clustering algorithm is the fuzzy $c$-means clustering algorithm \cite{bezdek-84}. With $K$ clusters, the $c$-means algorithm minimizes the generalized least-squared errors functional
\[
	J_{m}(U, v)=\sum_{i=1}^{n}\sum_{k=1}^{K}(u_{ik})^{m}||\mathbf{V}_{i}-v_{k}||^{2}, 
\]
where $m$ is a predefined weighting component, $U=[u_{ik}]$ is a fuzzy $c$-partition of a matrix $\mathbf{V}$, and $v=(v_{1},\dots,v_{K})$ is a vector of centers. Complete fuzzy clustering occurs when each unit has equal membership for all clusters (i.e., each membership coefficient is $1/K$), and a hard clustering occurs when each unit has a membership coefficient of 1 for one of the clusters. 

We consider several procedures to match units when using fuzzy clustering. First, we modify VM such that we classify all units using fuzzy clustering on $\text{logit}(\hat{R}_{t,t'}(\mathbf{X}))$. Because fuzzy clustering does not explicitly assign a cluster to any unit, we assign a unit to cluster $k\in K$ if its membership coefficient for that cluster is at least $1/K$. Units assigned to multiple clusters can be matched to other units that share at least one cluster with them. Second, we examined matching on the Mahalanobis distance of $\left\{\text{logit}(\hat{r}(t,\mathbf{X})),\text{logit}(\hat{r}(t',\mathbf{X}))\right\}$, within fuzzy clusters of $\text{logit}(\hat{R}_{t,t'}(\mathbf{X}))$. Third, we examined the effect of a caliper on these matching procedures. 

The fuzzy clustering procedures have similar steps 1, 2, and 4 as VM. Step 3 is modified to include fuzzy clustering based on the Mahalanobis distance. For a reference treatment $t\in\mathcal{W}$, these procedures can be summarized as follows:
\begin{enumerate}
	\item Estimate $R(\mathbf{X}_{i})$, $i=1,\dots,n$ using a multinomial logistic regression model. 
	\item Drop units outside the common support (e.g., those with $E_{i}=0$), and re-fit the model once. 
	\item For all $t'\neq t$:
	\begin{enumerate}
		\item Partition all units using fuzzy clustering on $\text{logit}(\hat{R}_{t,t'}(\mathbf{X}))$. A unit is part of cluster $k$ if its membership coefficient is at least $1/K$.  
		\item Within each strata $k\in \{1,\dots,K\}$, match those receiving $t$ to those receiving $t'$ on either the linear GPS (VMF), or the Mahalanobis distance of $\left\{\text{logit}(\hat{r}(t,\mathbf{X})),\text{logit}(\hat{r}(t',\mathbf{X}))\right\}$ with replacement and a caliper of 0.5 standard deviations (FM). FMnc performs the matching on the Mahalanobis distance of $\left\{\text{logit}(\hat{r}(t,\mathbf{X})),\text{logit}(\hat{r}(t',\mathbf{X}))\right\}$ without a caliper.  
	\end{enumerate} 
	\item Units receiving $t$ who were matched to units receiving all treatments $w\neq t$, along with their matches receiving the other treatments, comprise the final matched cohort. 
\end{enumerate}

We summarize and explicate each of the matching algorithms in Table~\ref{matching.methods}. All VM-related algorithms use the linear GPS as the distance measure, whereas clustering is performed using either $k$-means (VM, VM2, VMnc, VMnr) or fuzzy (VMF) clustering. The second class of algorithms used an initial clustering step using either $k$-means clustering (KM, KMnc) or fuzzy clustering (FM, FMnc) followed by matching on the Mahalanobis distance of $\left\{\text{logit}(\hat{r}(t,\mathbf{X})),\text{logit}(\hat{r}(t',\mathbf{X}))\right\}$ as the distance measure within clusters. The last class of methods, including LGPSM, LGPSMnc, and COVnc, are basic matching algorithms that do not use any initial clustering. 

\begin{table}[!htbp]
\centering
\caption{List of matching algorithms used}
\begin{tabular}{l c l l l}
\toprule
Label && Distance & Caliper (Y/N) & Clustering\\
\midrule
VM\cite{lopez-15} && linear GPS & Y & $k$-means\\
VM2 && linear GPS & Y & $k$-means\\
VMnc && linear GPS & N & $k$-means\\
VMnr && linear GPS & Y & $k$-means\\
VMF && linear GPS & Y & fuzzy\\
KM && Mahalanobis $\text{logit}(\hat{r}(t,\mathbf{X}),\hat{r}(t',\mathbf{X}))$ & Y & $k$-means\\
KMnc && Mahalanobis $\text{logit}(\hat{r}(t,\mathbf{X}),\hat{r}(t',\mathbf{X}))$ & N & $k$-means\\
FM && Mahalanobis $\text{logit}(\hat{r}(t,\mathbf{X}),\hat{r}(t',\mathbf{X}))$ & Y & fuzzy\\
FMnc && Mahalanobis $\text{logit}(\hat{r}(t,\mathbf{X}),\hat{r}(t',\mathbf{X}))$ & N & fuzzy\\
LGPSM && Mahalanobis logit GPS vector & Y & NA\\
LGPSMnc && Mahalanobis logit GPS vector & N & NA\\
COVnc && Mahalanobis covariates & N & NA\\
\bottomrule
\end{tabular}
\label{matching.methods}
\end{table}

\section{Simulations}

\subsection{Evaluating balance of matched sets by simulation}

To compare the performance of the different matching algorithms described in Section~\ref{sec:matching}, we performed extensive simulations. Simulation configurations were either known to the investigator or can be estimated from the data. A $P$-dimensional $\mathbf{X}$ was generated for $n=n_{1}+\cdots+n_{Z}$ units receiving one of $Z\in\{3,5,10\}$ treatments, $\mathcal{W}=\{1,\dots,Z\}$. For $Z=3$, we generated sample sizes such that $n_{2}=\gamma n_{1}$ and $n_{3}=\gamma^{2}n_{1}$. For $Z=5$, we generated similar sample sizes for $n_{1}$, $n_{2}$, and $n_{3}$ as for $Z=3$, and we set $n_{4}=n_{2}$ and $n_{5}=n_{3}$. For $Z=10$, the treatment group sample sizes for $n_{1},\dots,n_{5}$ were the same as for $Z=5$, and the sizes of treatment groups 6-10 were $n_{i+5}=n_{i}$, $i=1,\dots,5$. 

The values of $\mathbf{X}$ were generated from multivariate skew-$t$ distributions such that 
\[
	\mathbf{X}_{i}\mid \{W_{i}=w\}\sim \text{Skew-}t_{df,P}(\boldsymbol\mu_{w}, \boldsymbol\Sigma_{w}, \boldsymbol\eta).
\] 
For $Z\in\{3,5\}$, $\boldsymbol\mu_{w}=\text{vec}(1_{P}\otimes b_{w})$, where $1_{P}$ is a $P\times1$ vector of 1s, and $b_{w}$ is the $Z\times1$ vector such that the $w$ value is equal to $b$ and the rest are zeros. In addition, the covariance matrices $\boldsymbol\Sigma_{1}$, $\boldsymbol\Sigma_{2}$, $\boldsymbol\Sigma_{3}$, $\boldsymbol\Sigma_{4}$, and $\boldsymbol\Sigma_{5}$ have respective diagonal entries of 1, $\sigma_{2}^{2}$, $\sigma_{3}^{2}$, $\sigma_{2}^{2}$, and $\sigma_{3}^{2}$, and $\lambda$ elsewhere. For $Z=10$, $b_{w}$ is a $10\times1$ vector such that the $w$ value is equal to $b$ and the rest are zeros, and the covariates matrices $\boldsymbol\Sigma_{w}$, $w=1,\dots,10$ have respective diagnonal entries of 1, $\sigma_{2}^{2}$, $\sigma_{3}^{2}$, $\sigma_{2}^{2}$, $\sigma_{3}^{2}$, 1, $\sigma_{2}^{2}$, $\sigma_{3}^{2}$, $\sigma_{2}^{2}$, and $\sigma_{3}^{2}$, and $\lambda$ elsewhere. This was done in order to reduce the running time when dealing with a large number of treatments.  

The simulation design assumes a regular assignment mechanism that depends on the factors listed in Table~\ref{sim.factors}, resulting in a $2^{4}\times3^{4}\times5$ factorial design for $Z\in\{3,5\}$, and a $2^{7}\times5$ factorial design for $Z=10$. For each configuration, 100 replications were produced. For $Z\in\{3,5\}$ we discarded configurations where $P=20$ and $n_{1}=600$, and for $Z\in\{3,5,10\}$ when $P=20$ and $b=1$, because of the small number of units that can be matched across all treatment arms. After discarding these configurations, there were 5,184 simulation configurations for $Z\in\{3,5\}$, and 576 simulation configurations for $Z=10$. The simulation design is an extension to the ones described in Lopez and Gutman (2017) \cite{lopez-15}. The main differences between the simulation designs are the inclusion of more covariates as well as more possible interventions. 

\begin{table}[!htbp]
\centering
\caption{Simulation factors}
\begin{tabular}{l c l c l}
\toprule
Factor && $Z\in\{3,5\}$ levels && $Z=10$ levels\\
\midrule
$n_{1}$ && $\{600, 1200\}$ && 900\\
$\gamma$ && $\{1,2\}$ && $\{1,2\}$\\
$b$ && $\{0, 0.25,0.50,0.75,1\}$ && $\{0,0.25,0.50,0.75,1\}$\\
$\lambda$ && $\{0, 0.25\}$ && $\{0, 0.25\}$\\
$\sigma_{2}^{2}$ && $\{0.5, 1, 2\}$ && $\{1, 2\}$\\
$\sigma_{3}^{2}$ && $\{0.5, 1, 2\}$ && $\{1, 2\}$\\
$\eta$ && $\{-3.5, 0, 3.5\}$ && $\{0, 3.5\}$\\
$df$ && $\{7, \infty\}$ && $\{7,\infty\}$\\
$P$ && $\{5, 10, 20\}$ && $\{10,20\}$\\
\bottomrule
\end{tabular}
\label{sim.factors}
\end{table}

To estimate the GPS model we used two different multinomial logistic regression models. The first model includes all of the covariates, $\mathbf{X}_i$, as explanatory variables. The second model includes all of the covariates as well as the square of all covariates, $\mathbf{X}_i^{2}$. Throughout, we will refer to the first model as the first-order GPS model and to the second model as the second-order GPS model. All of the simulations were conducted using R Studio software \cite{RSoftware}, and all matching algorithms were implemented using the Matching package \cite{sekhon-11}. 

\subsection{Simulation metrics}

Let $\psi_{iw}$ be the number of times that unit $i$ serves as a match to other units in treatment group $w$, and let $n_{wm}$ be the number of units from treatment group $w$ in the matched sample, including units that are used as a match more than once. The weighted mean of covariate $p$, $p=1,\dots,P$, at treatment $w$, is defined as $\bar{X}_{pw}$, such that
\[
	\bar{X}_{pw}=\frac{1}{n_{wm}}\sum_{i=1}^{n}X_{pi}T_{iw}\psi_{iw}.
\]

We define the standardized bias at each covariate $p$ for each pair of treatments as
\[
	SB_{pjk}=\frac{\bar{X}_{pj}-\bar{X}_{pk}}{\delta_{pt}}, 
\]
where $\delta_{pt}$ is the standard deviation of $\mathbf{X}_{p}$ in the full sample among units receiving the reference treatment $W=t$. As in McCaffrey et al. (2013) \cite{mccaffrey-13} and Lopez and Gutman (2017) \cite{lopez-15}, we extract the maximum absolute standardized pairwise bias at each covariate, 
\[
	Max2SB_{p}=\max\left(|SB_{p12}|, |SB_{p13}|, |SB_{p23}|, \dots\right).
\]
This metric reflects the largest discrepancy in estimated means between any two treatment groups for covariate $p$. McCaffrey et al. (2013) \cite{mccaffrey-13} advocated a cutoff of 0.20, but maintained that larger cutoffs may be appropriate for different applications. 

For matching algorithms with a caliper, a second metric that is used to measure matching performance is the proportion of units from the eligible population with $W=1$ that were included in the final matched set, $Prop.Matched$. When estimating $\tau_{jk}^{t}$, scenarios with $Prop.Matched\approx1$ and low $Max2SB_{p}$ for all $p$ are optimal, because most of the units in the population of interest are retained, and the covariates' distributions are similar on average across treatment groups. By design, matching algorithms without a caliper have $Prop.Matched=1$. 

A third metric that is used to measure balance among matched sets is the ratio of the variances of covariates between each treatment group \cite{stuart-10, rubin-07}. In the multiple treatment setting, we extract the maximum absolute pairwise difference between variances at each covariate, measured on the natural logarithm scale, 
\[
	Max2log_{p}=\max\left(\left\lvert\log\left(\frac{Var(X_{p1})}{Var(X_{p2})}\right)\right\rvert, \left\lvert\log\left(\frac{Var(X_{p1})}{Var(X_{p3})}\right)\right\rvert, \left\lvert\log\left(\frac{Var(X_{p2})}{Var(X_{p3})}\right)\right\rvert,\dots\right), 
\]
where
\[
	Var(X_{pw}) = \frac{\sum_{i=1}^{n}(X_{pi}-\bar{X}_{pw})^{2}T_{iw}\psi_{iw}}{n_{wm}}.
\]
This metric summarizes the balance of the second moments of covariate $p$ among all treatment groups. For binary treatment, Rubin \cite{rubin-07} classified the ratio of variances between treatment groups that is in $[4/5,5/4]$ as "good," variance ratios in $[1/2,4/5)$ or $(5/4,2]$ as "of concern," and variance ratios below $1/2$ or above 2 as "bad." These thresholds correspond to cutoffs  of 0.22 and 0.69 on the absolute log scale. 

For each configuration, at each replication, we calculate $Max2SB_{p}$ and $Max2log_{p}$ for all $p$, and $Prop.Matched$. We summarize $Max2SB_{p}$ for all $P$ covariates using $MaxMax2SB\equiv \max_{p=1,\dots,P}(Max2SB_{p})$ and $\overline{Max2SB}\equiv(1/P)\sum_{p=1}^{P}Max2SB_{p}$, and we summarized $Max2log_{p}$ for all $P$ covariates using $MaxMax2log\equiv \max_{p=1,\dots,P}(Max2log_{p})$. Lastly, we averaged $MaxMax2SB$, $\overline{Max2SB}$, and $MaxMax2log$ across the 100 replications. 

We provide results for $MaxMax2SB$ because it represents the largest imbalance among all covariates, and it is not as affected as $\overline{Max2SB}$ by increasing the number of covariates with no biases. Results for $\overline{Max2SB}$ show similar trends and are available in the Appendix. 

\subsection{Simulation results, $Z=3$}

\begin{figure}[!t]
\centering
\includegraphics[scale=0.5]{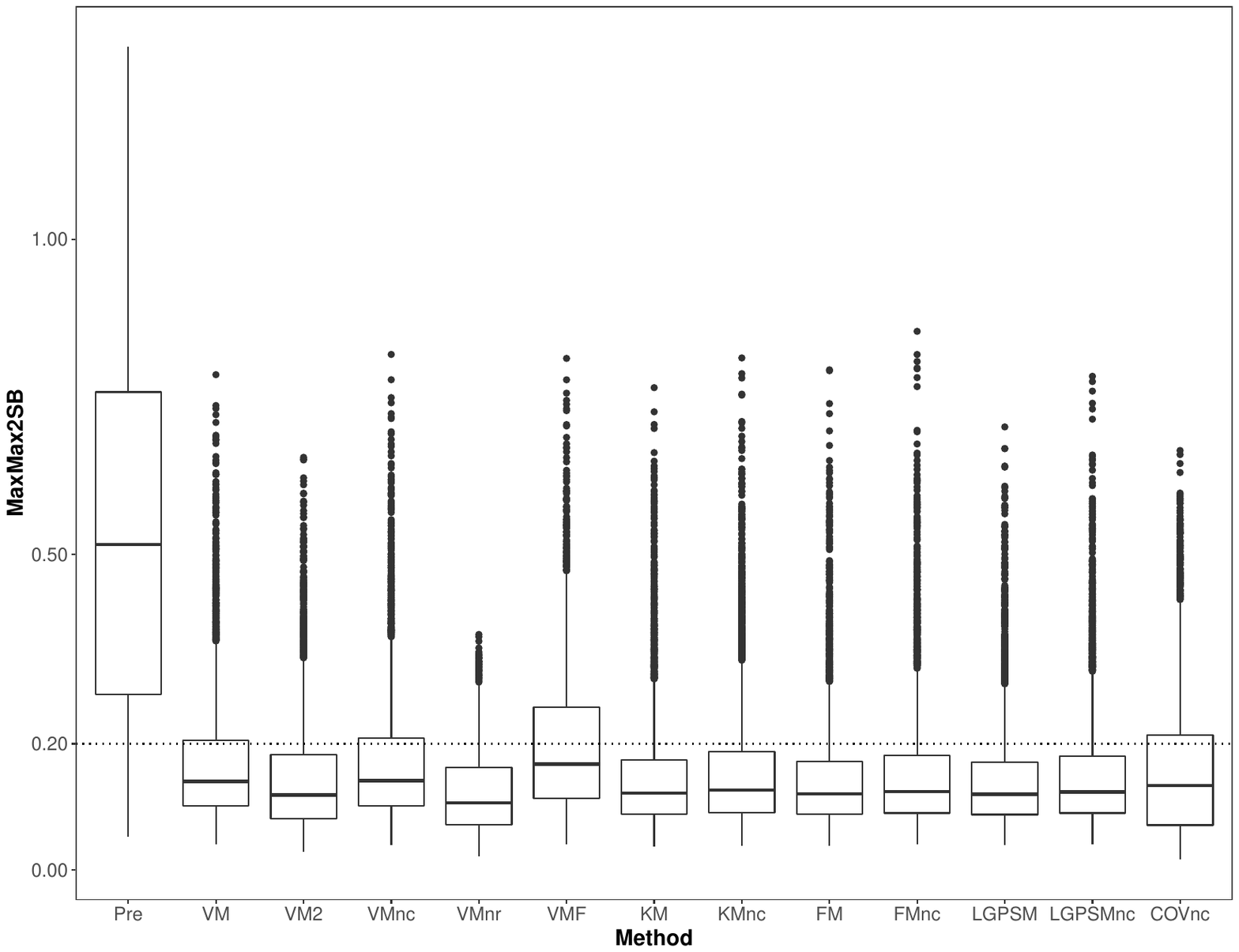}
\caption{$MaxMax2SB$ for pre-matched cohort and by matching algorithms, $Z=3$}
\label{max3}
\end{figure}

Figure~\ref{max3} shows boxplots of $MaxMax2SB$ across simulation factors for $Z=3$, for the pre-matched cohort of eligible units and the ten matching algorithms, when using a first-order GPS model. Each point in the boxplots represents the maximal bias at one factors' configuration. The best-performing method is VMnr with only 12\% of the configurations above 0.2. It is followed by LGPSM and FM, with 17\% and 18\% of the configurations above 0.2, respectively. COVnc has the worst performance, with $MaxMax2SB$ exceeding 0.2 in 38\% of configurations.

When using a second-order GPS model, the best-performing method with second-order GPS is VMnr with 38\% of the configurations above 0.2. It is followed by COVnc, LGPSM, and FM, with $MaxMax2SB$ exceeding 0.2 in 39\%, 45\%, and 45\% of configurations, respectively. 

\begin{figure}[!t]
\centering
\includegraphics[scale=0.5]{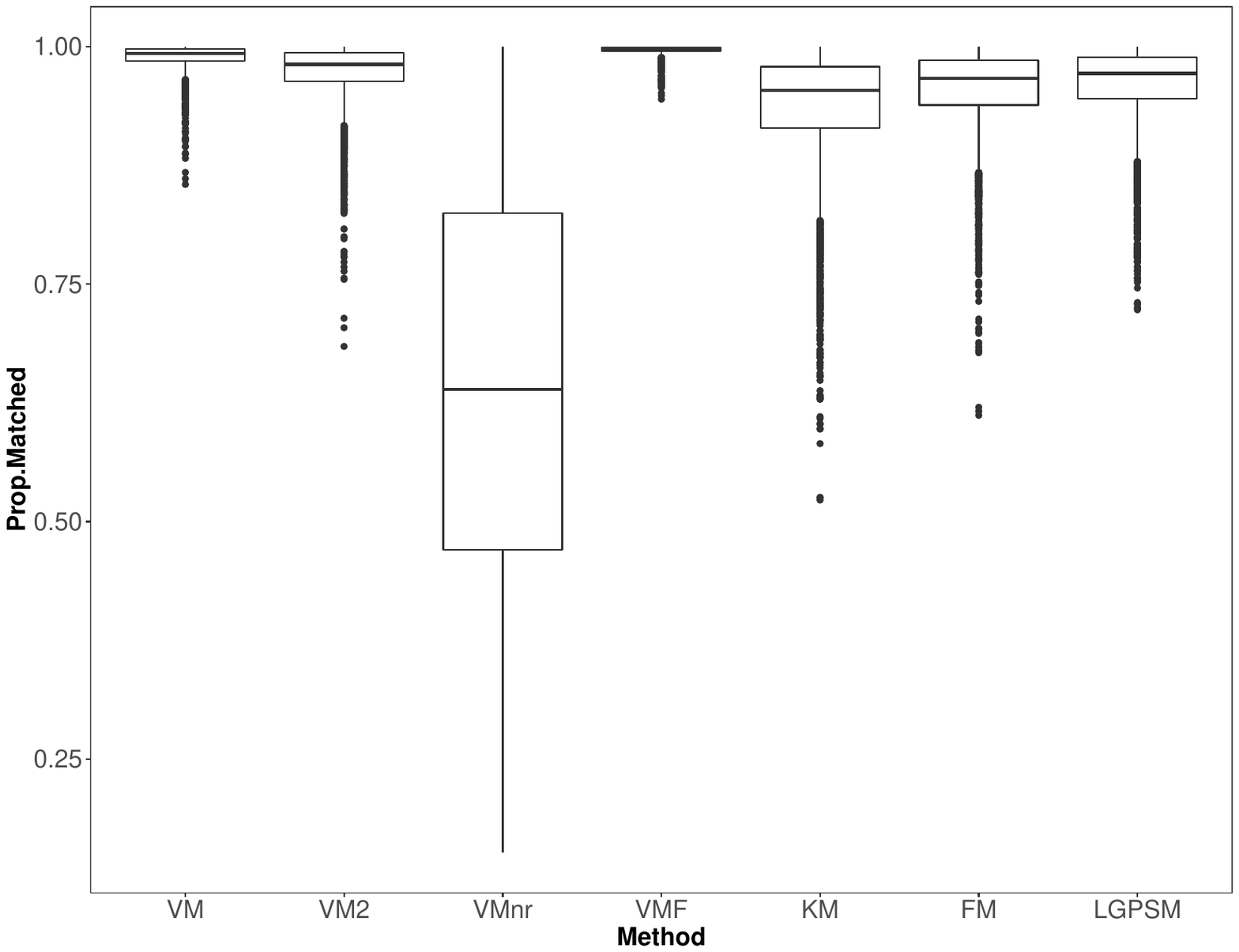}
\caption{$Prop.Matched$ for algorithms that used a caliper, $Z=3$}
\label{prop3}
\end{figure}

Figure~\ref{prop3} depicts $Prop.Matched$ across configurations for matching algorithms with a caliper. VMF has the highest median $Prop.Matched$ with 99.8\% of the reference group units being matched on average. VM and VM2 have the second and third highest median $Prop.Matched$, with 99.3\% and 98.1\% of the reference group units matched on average, respectively. VMnr has the lowest median $Prop.Matched$, with only 64\% of the reference group units matched on average. Thus, while VMnr generally yields the lowest bias in the matched cohort, it is at the expense of generalizability, because the matched cohort is less representative of the original sample. 

To identify factors with the largest influence on the performance of each matching algorithm, we rank them by their MSE for $MaxMax2SB$ (as in \cite{lopez-15, rubin-79, cangul-09}). For a first-order GPS model, initial covariate bias, $b$, explains the largest portion of variation in $MaxMax2SB$, accounting for at least 35\% of the variability for each algorithm (data not shown). The number of covariates, $P$, explains the second largest portion of variation in $MaxMax2SB$, accounting for at least 20\% of the variability for most of the procedures, except for VMnr. Other influential factors include the interaction between $b$ and $P$, and the ratio of units with $W = 2$ to units with $W = 1$ ($\gamma$), though these do not account for as much variability in $MaxMax2SB$ as $b$ or $P$. Similar trends are observed for the second-order GPS (data not shown).

Table~\ref{bP3table} shows $MaxMax2SB$ based on different levels of $b$, $P$, and the two GPS models. In settings with low covariate bias, all ten matching algorithms appear to balance the covariates properly, regardless of the number of total covariates or GPS model order. As $b$ increases, $MaxMax2SB$ increases for all algorithms, though it increases more rapidly for matching on the Mahalanobis distance of the covariates (COVnc) when $P\geq 10$. For $P=5$, $MaxMax2SB$ increases at roughly the same rate for all matching algorithms as $b$ increases. Except for COVnc, the median $MaxMax2SB$ for each algorithm is higher across $b$ and $P$ when using a second-order GPS model. Among the methods that retain all of the units in $W=1$, LGPSMnc has the lowest overall bias for most of the configurations and it is comparable to VMnr.

\begin{table}[]
\footnotesize
\centering
\caption{Median $MaxMax2SB$ across levels of $b$, $P$, and GPS model order, $Z=3$ (smallest $MaxMax2SB$ in italics)}
\begin{tabular}{c c c c c c c c c c c c c}
\multicolumn{13}{c}{First-order GPS model}\\
\toprule
& \multicolumn{12}{c}{$P=5$}\\
\cline{2-13}
$b$ & VM & VM2 & VMnc & VMnr & VMF & KM & KMnc & FM & FMnc & LGPSM & LGPSMnc & COVnc\\
\midrule
0.00 & 0.09 & 0.07 & 0.09 & 0.06 & 0.09 & 0.08 & 0.08 & 0.08 & 0.08 & 0.08 & 0.08 & \textit{0.04}\\
0.25 & 0.11 & 0.09 & 0.11 & 0.08 & 0.11 & 0.10 & 0.10 & 0.10 & 0.10 & 0.10 & 0.10 & \textit{0.06}\\
0.50 & 0.13 & 0.11 & 0.13 & 0.11 & 0.13 & 0.11 & 0.12 & 0.11 & 0.11 & 0.11 & 0.11 & \textit{0.10}\\
0.75 & 0.17 & 0.15 & 0.17 & 0.15 & 0.16 & \textit{0.13} & 0.14 & \textit{0.13} & \textit{0.13} & \textit{0.13} & \textit{0.13} & 0.15\\
1.00 & 0.22 & 0.20 & 0.23 & 0.21 & 0.21 & 0.17 & 0.18 & \textit{0.16} & 0.17 & \textit{0.16} & 0.17 & 0.21\\ 
& \multicolumn{12}{c}{$P=10$}\\
\cline{2-13}
$b$ & VM & VM2 & VMnc & VMnr & VMF & KM & KMnc & FM & FMnc & LGPSM & LGPSMnc & COVnc\\
\midrule
0.00 & 0.08 & 0.06 & 0.08 & \textit{0.05} & 0.08 & 0.08 & 0.08 & 0.07 & 0.07 & 0.07 & 0.07 & 0.07\\
0.25 & 0.09 & 0.07 & 0.09 & \textit{0.06} & 0.09 & 0.09 & 0.09 & 0.09 & 0.09 & 0.09 & 0.09 & 0.13\\ 
0.50 & 0.12 & 0.10 & 0.12 & \textit{0.09} & 0.12 & 0.11 & 0.11 & 0.11 & 0.11 & 0.11 & 0.11 & 0.21\\
0.75 & 0.18 & 0.16 & 0.18 & \textit{0.13} & 0.17 & 0.15 & 0.16 & 0.15 & 0.16 & 0.15 & 0.16 & 0.31\\
1.00 & 0.27 & 0.24 & 0.27 & \textit{0.18} & 0.26 & 0.24 & 0.25 & 0.23 & 0.24 & 0.22 & 0.23 & 0.41\\ 
& \multicolumn{12}{c}{$P=20$}\\
\cline{2-13}
$b$ & VM & VM2 & VMnc & VMnr & VMF & KM & KMnc & FM & FMnc & LGPSM & LGPSMnc & COVnc\\
\midrule
0.00 & 0.13 & 0.09 & 0.12 & \textit{0.08} & 0.12 & 0.12 & 0.12 & 0.11 & 0.12 & 0.11 & 0.12 & 0.10\\
0.25 & 0.15 & 0.12 & 0.15 &\textit{0.10} & 0.15 & 0.14 & 0.14 & 0.14 & 0.14 & 0.13 & 0.14 & 0.19\\
0.50 & 0.24 & 0.21 & 0.25 & \textit{0.16} & 0.24 & 0.22 & 0.24 & 0.21 & 0.23 & 0.21 & 0.22 & 0.32\\
0.75 & 0.43 & 0.38 & 0.44 & \textit{0.26} & 0.43 & 0.39 & 0.42 & 0.39 & 0.41 & 0.37 & 0.40 & 0.46\\
\bottomrule\\
\multicolumn{13}{c}{Second-order GPS model}\\
\toprule
& \multicolumn{12}{c}{$P=5$}\\
\cline{2-13}
$b$ & VM & VM2 & VMnc & VMnr & VMF & KM & KMnc & FM & FMnc & LGPSM & LGPSMnc & COVnc\\
\midrule
0.00 & 0.12 & 0.10 & 0.13 & 0.09 & 0.14 & 0.11 & 0.11 & 0.11 & 0.11 & 0.11 & 0.11 & \textit{0.04}\\
0.25 & 0.13 & 0.11 & 0.13 & 0.11 & 0.17 & 0.11 & 0.12 & 0.11 & 0.12 & 0.11 & 0.12 & \textit{0.06}\\
0.50 & 0.15 & 0.13 & 0.15 & 0.13 & 0.20 & 0.12 & 0.13 & 0.12 & 0.13 & 0.12 & 0.12 & \textit{0.10}\\
0.75 & 0.18 & 0.17 & 0.19 & 0.18 & 0.26 & 0.15 & 0.16 & \textit{0.14} & 0.15 & 0.15 & 0.15 & 0.15\\
1.00 & 0.24 & 0.23 & 0.24 & 0.24 & 0.34 & 0.19 & 0.21 & \textit{0.18} & 0.20 & \textit{0.18} & 0.19 & 0.21\\ 
& \multicolumn{12}{c}{$P=10$}\\
\cline{2-13}
$b$ & VM & VM2 & VMnc & VMnr & VMF & KM & KMnc & FM & FMnc & LGPSM & LGPSMnc & COVnc\\
\midrule
0.00 & 0.16 & 0.14 & 0.17 & 0.12 & 0.18 & 0.15 & 0.16 & 0.15 & 0.15 & 0.15 & 0.15 & \textit{0.07}\\
0.25 & 0.18 & 0.15 & 0.18 & 0.14 & 0.20 & 0.16 & 0.17 & 0.16 & 0.17 & 0.16 & 0.17 & \textit{0.13}\\ 
0.50 & 0.23 & 0.21 & 0.24 & \textit{0.17} & 0.26 & 0.21 & 0.22 & 0.21 & 0.22 & 0.20 & 0.21 & 0.22\\
0.75 & 0.33 & 0.30 & 0.34 & \textit{0.24} & 0.36 & 0.29 & 0.31 & 0.29 & 0.30 & 0.27 & 0.29 & 0.32\\
1.00 & 0.47 & 0.42 & 0.49 & \textit{0.32} & 0.50 & 0.42 & 0.46 & 0.41 & 0.44 & 0.39 & 0.43 & 0.42\\ 
& \multicolumn{12}{c}{$P=20$}\\
\cline{2-13}
$b$ & VM & VM2 & VMnc & VMnr & VMF & KM & KMnc & FM & FMnc & LGPSM & LGPSMnc & COVnc\\
\midrule
0.00 & 0.19 & 0.16 & 0.19 & 0.14 & 0.19 & 0.19 & 0.19 & 0.19 & 0.19 & 0.18 & 0.19 & \textit{0.10}\\
0.25 & 0.23 & 0.21 & 0.24 & \textit{0.15} & 0.25 & 0.22 & 0.23 & 0.22 & 0.23 & 0.22 & 0.22 & 0.19\\
0.50 & 0.36 & 0.32 & 0.36 & \textit{0.21} & 0.37 & 0.34 & 0.36 & 0.34 & 0.35 & 0.33 & 0.34 & 0.32\\
0.75 & 0.57 & 0.50 & 0.58 & \textit{0.31} & 0.57 & 0.53 & 0.58 & 0.53 & 0.57 & 0.50 & 0.54 & 0.46\\
\bottomrule
\end{tabular}
\label{bP3table}
\end{table}

\begin{figure}[!t]
\centering
\includegraphics[scale=0.5]{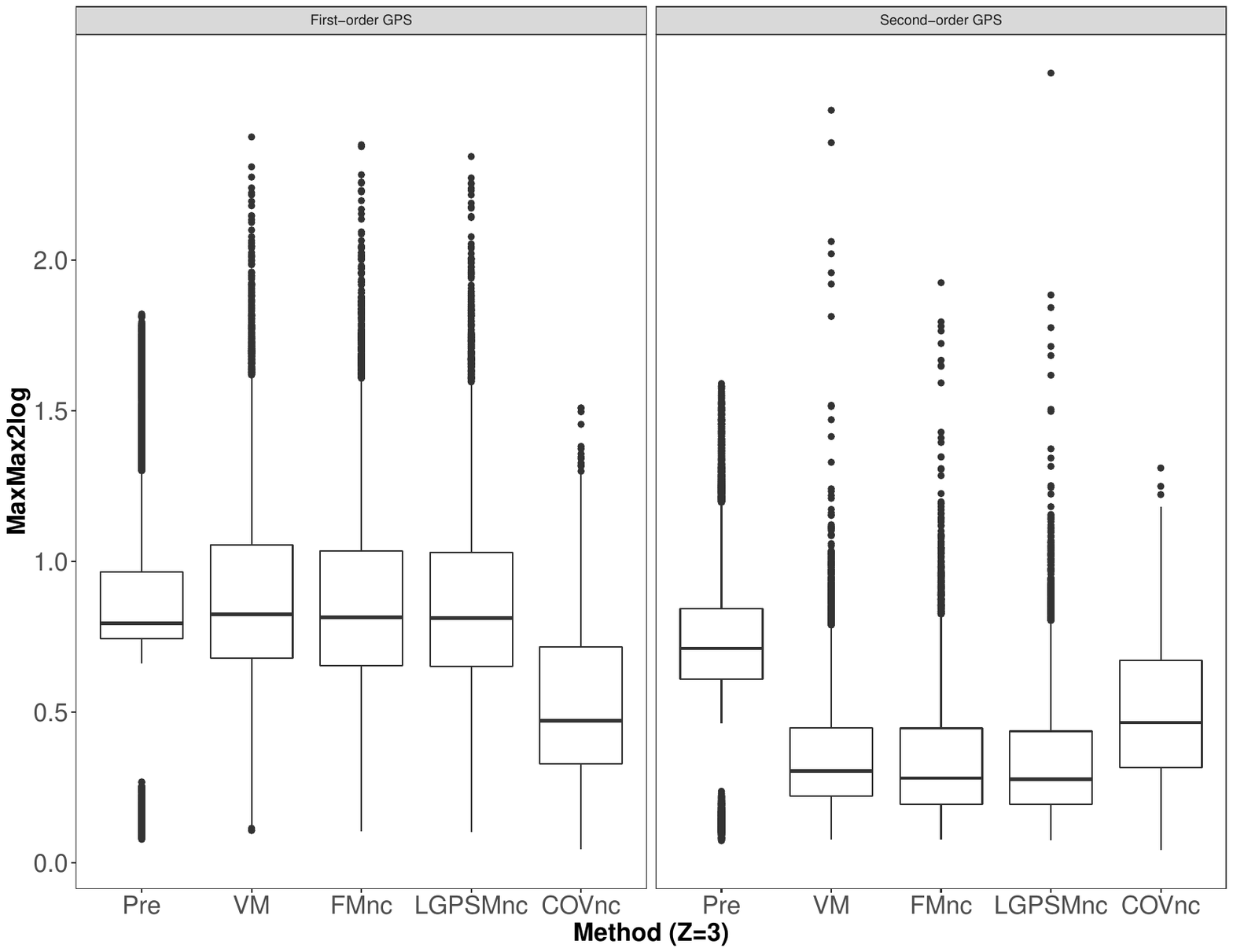}
\caption{$MaxMax2log$ for pre-matched cohort across matching algorithms, $Z=3$}
\label{balance3.log}
\end{figure}

The $MaxMax2log$ for sets obtained using either VM, FMnc, LGPSMnc, or COVnc are depicted in Figure~\ref{balance3.log}. For each algorithm, the median $MaxMax2log$ is significantly larger than the 0.22 cutoff when using the first-order GPS. When including the square of the covariate terms in the GPS model, the $MaxMax2log$ is lower for each algorithm that uses the GPS for matching. VM, FMnc, and LGPSMnc yield $MaxMax2log$ below 0.22 for 24\%, 34\%, and 34\% of configurations, and below 0.69 for 93\%, 92\%, and 93\% of configurations, respectively.

The most influential simulation factors for $MaxMax2log$ were $\sigma_{2}^{2}$ and $\sigma_{3}^{2}$ for both GPS models. Table~\ref{s2s3table3} in the Appendix shows the median $MaxMax2log$ based on different levels of $\sigma_{2}^{2}$, $\sigma_{3}^{2}$, and the two GPS models. For VM, FMnc, and LGPSMnc, $MaxMax2log$ is smallest when $\sigma_{2}^{2} = \sigma_{3}^{2} = 1$. This  configuration implies that the initial variances are equal for all three treatment groups across all of the covariates. $MaxMax2log$ is largest when the distance between $\sigma_{2}^{2}$ and $\sigma_{3}^{2}$ is largest, such as when $\sigma_{2}^{2}=2.0$ and $\sigma_{3}^{2}=0.5$ or vice-versa. These trends are observed for both GPS models, though $MaxMax2log$ was smaller overall for each level of $\sigma_{2}^{2}$ and $\sigma_{3}^{2}$ when using a second-order GPS model. Because COVnc only uses the GPS model in defining the region of common support, it is only slightly impacted by the choice of GPS model. Thus, the smaller bias in the covariates when using COVnc is at the expense of larger variance ratios.

\subsection{Simulation results, $Z=5$}

\begin{figure}[!t]
\centering
\includegraphics[scale=0.5]{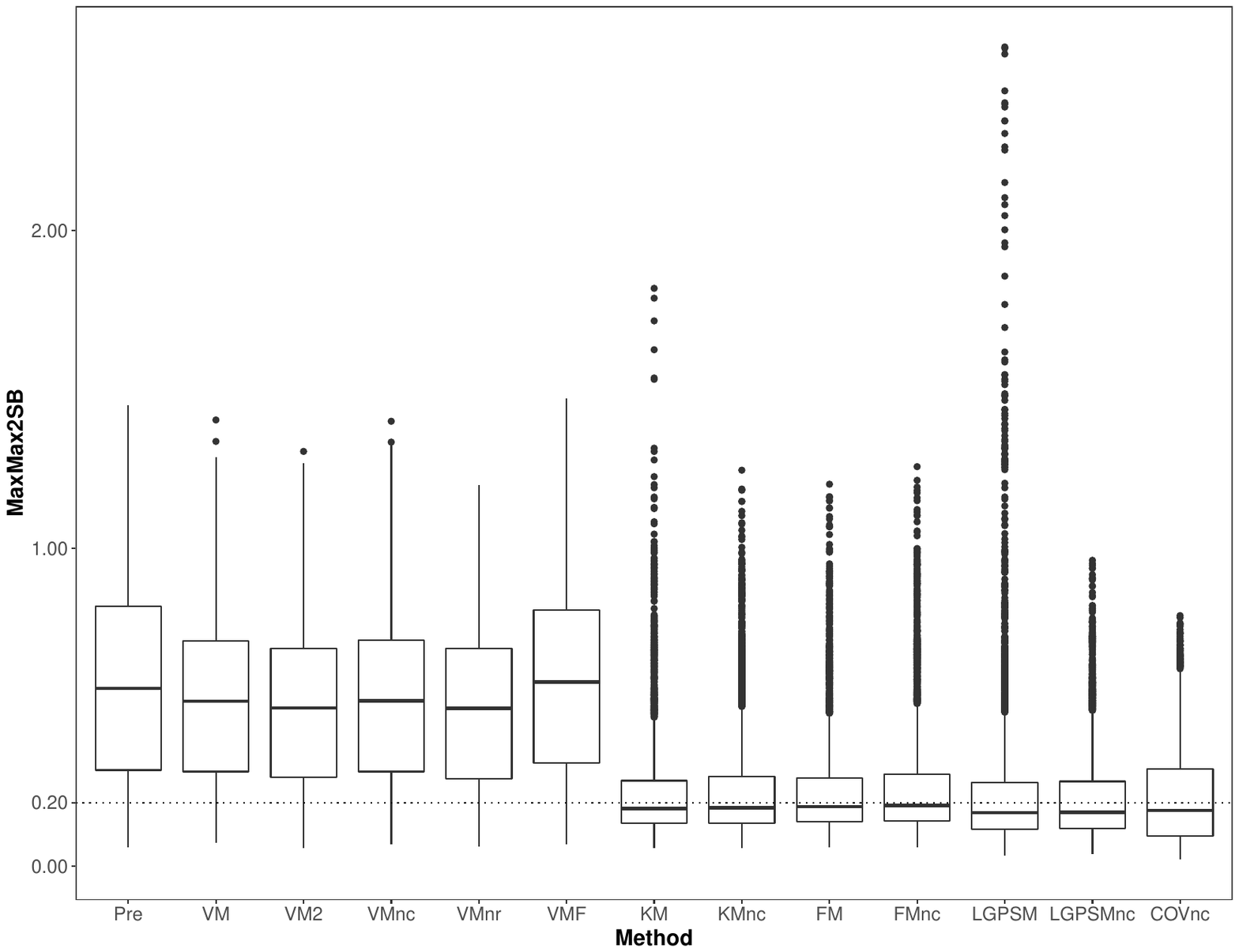}
\caption{$MaxMax2SB$ for pre-matched cohort and by matching algorithms, $Z=5$}
\label{max5}
\end{figure}

Figure~\ref{max5} shows boxplots of $MaxMax2SB$ across simulation factors for $Z=5$, for each of the ten matching algorithms and in the pre-matched cohort of eligible units. VM exceeds 0.2 for the majority of the configurations. The algorithms without a caliper (KMnc, FMnc, LGPSMnc, COVnc) perform more favorably for five treatments, with the majority of the configurations yielding a $MaxMax2SB$ below 0.20. KM and FM also perform favorably, with 57\% and 55\% of configurations yielding a $MaxMax2SB$ below 0.20, respectively. 

Similar to $Z=3$, when using a second-order GPS model, the best-performing methods are similar to those when using a first-order GPS model, though $MaxMax2SB$ is generally higher (data not shown). The algorithms without a caliper which use a first-order GPS model (KMnc, FMnc, LGPSMnc) yield a $MaxMax2SB$ below 0.20 for 41\%, 41\%, and 44\% of configurations, respectively. KM and FM perform similarly, with 47\% and 45\% of configurations yielding a $MaxMax2SB$ below 0.20, respectively. COVnc performs the best when using a second-order GPS model, with 56\% of configurations yielding a $MaxMax2SB$ below 0.20. 

\begin{figure}[!t]
\centering
\includegraphics[scale=0.5]{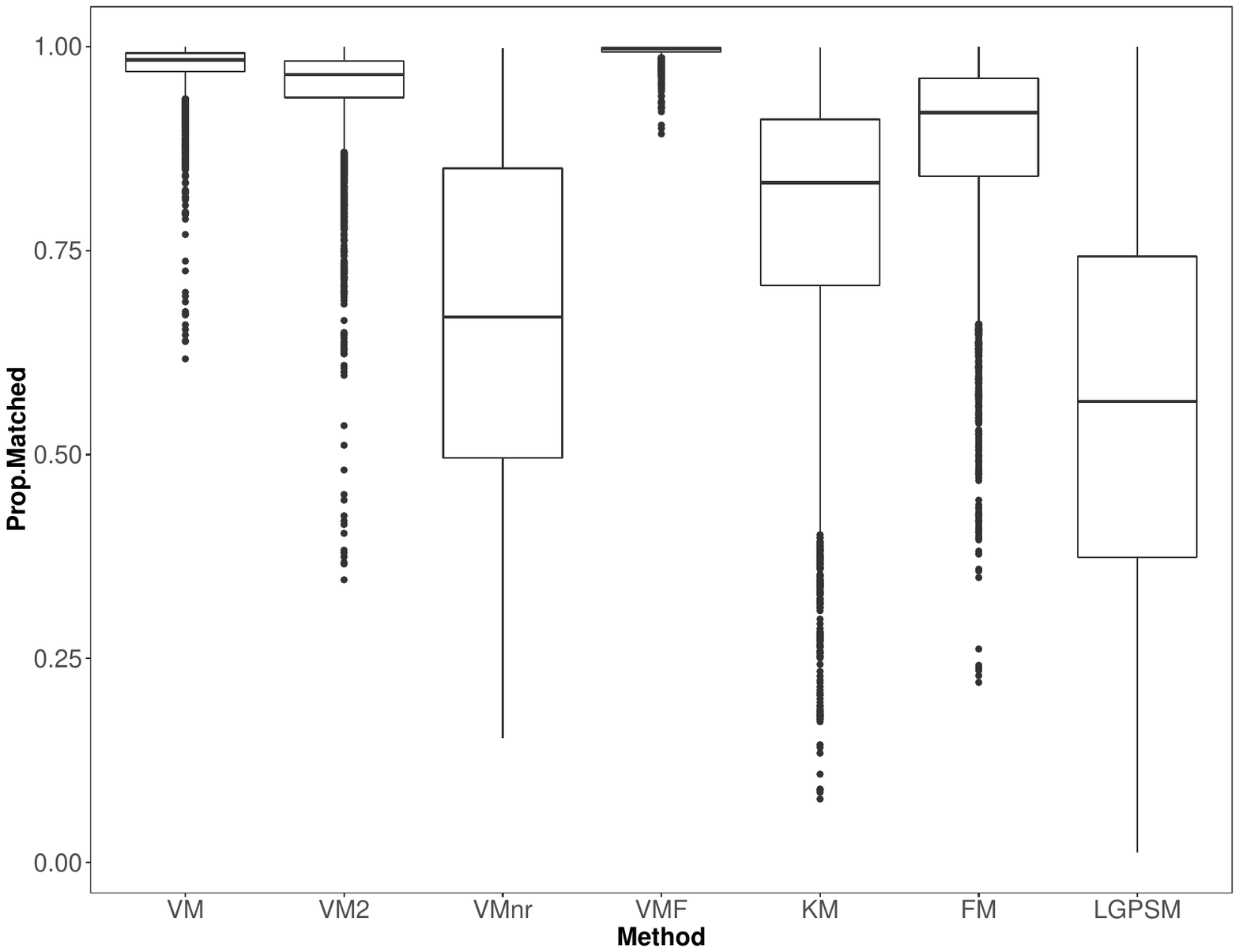}
\caption{$Prop.Matched$ for algorithms that used a caliper, $Z=5$}
\label{prop5}
\end{figure}

The only matching algorithms with a caliper that yield a median $Prop.Matched$ exceeding 0.90 are the VM with replacement algorithms and FM (Figure~\ref{prop5}). FM matches more than 75\% of reference group units in 68\% of configurations, and VM with-replacement algorithms match more than 75\% of the units in the reference group in 98\% of the configurations.

Similar to $Z=3$, initial covariate bias explains the highest portion of variability in $MaxMax2SB$, accounting for at least 25\% of the variability for each algorithm (data not shown). The number of covariates, $P$, explains the second largest portion of the variability in $MaxMax2SB$ for most of the algorithms. The ratio of treatment group sample sizes explains the third largest portion of the variability in $MaxMax2SB$. 

Table~\ref{bP5table} shows the median $MaxMax2SB$ for different levels of $b$, $P$, and the two GPS models. Overall trends for $b$, $P$, and GPS model are similar to those observed for $Z=3$. However, VM algorithms have the worse performance with increasing $b$. The medians $MaxMax2SB$ for these algorithms are larger than 0.20 for $b>0$ across all levels of $P$. Algorithms which match on the Mahalanobis distance of the logit GPS or the Mahalanobis distance of a subset of the logit GPS have medians $MaxMax2SB$ that exceed 0.20 for $b=1$ when $P=5$, and for $b\geq0.50$ when $P\in\{10,20\}$. 

\begin{table}[!htbp]
\footnotesize
\centering
\caption{Median $MaxMax2SB$ across levels of $b$, $P$, and GPS model order, $Z=5$ (smallest $MaxMax2SB$ in italics)}
\begin{tabular}{c c c c c c c c c c c c c}
\multicolumn{13}{c}{First-order GPS model}\\
\toprule
& \multicolumn{12}{c}{$P=5$}\\
\cline{2-13}
$b$ & VM & VM2 & VMnc & VMnr & VMF & KM & KMnc & FM & FMnc & LGPSM & LGPSMnc & COVnc\\
\midrule
0.00 & 0.14 & 0.12 & 0.14 & 0.11 & 0.14 & 0.11 & 0.10 & 0.11 & 0.11 & 0.08 & 0.08 & \textit{0.04}\\
0.25 & 0.33 & 0.31 & 0.33 & 0.31 & 0.33 & 0.13 & 0.13 & 0.14 & 0.14 & 0.10 & 0.10 & \textit{0.07}\\
0.50 & 0.54 & 0.53 & 0.54 & 0.53 & 0.56 & 0.15 & 0.15 & 0.17 & 0.17 & \textit{0.12} & 0.13 & \textit{0.12}\\
0.75 & 0.70 & 0.69 & 0.70 & 0.71 & 0.73 & 0.18 & 0.19 & 0.20 & 0.21 & \textit{0.14} & 0.18 & 0.17\\
1.00 & 0.82 & 0.81 & 0.83 & 0.86 & 0.86 & 0.24 & 0.25 & 0.25 & 0.26 & \textit{0.18} & 0.24 & 0.24\\ 
& \multicolumn{12}{c}{$P=10$}\\
\cline{2-13}
$b$ & VM & VM2 & VMnc & VMnr & VMF & KM & KMnc & FM & FMnc & LGPSM & LGPSMnc & COVnc\\
\midrule
0.00 & 0.17 & 0.13 & 0.17 & 0.13 & 0.16 & 0.14 & 0.14 & 0.14 & 0.14 & 0.14 & 0.12 & \textit{0.08}\\
0.25 & 0.35 & 0.33 & 0.35 & 0.32 & 0.36 & 0.16 & 0.16 & 0.17 & 0.16 & 0.17 & \textit{0.15} & \textit{0.15}\\ 
0.50 & 0.54 & 0.52 & 0.54 & 0.52 & 0.56 & 0.21 & 0.21 & 0.21 & 0.21 & 0.21 & \textit{0.20} & 0.24\\
0.75 & 0.69 & 0.66 & 0.70 & 0.67 & 0.71 & \textit{0.29} & 0.31 & 0.30 & 0.31 & 0.30 & \textit{0.29} & 0.35\\
1.00 & 0.85 & 0.81 & 0.86 & 0.79 & 0.84 & \textit{0.44} & 0.48 & 0.47 & 0.50 & 0.45 & \textit{0.44} & 0.47\\ 
& \multicolumn{12}{c}{$P=20$}\\
\cline{2-13}
$b$ & VM & VM2 & VMnc & VMnr & VMF & KM & KMnc & FM & FMnc & LGPSM & LGPSMnc & COVnc\\
\midrule
0.00 & 0.16 & 0.13 & 0.16 & 0.12 & 0.16 & 0.14 & 0.14 & 0.14 & 0.14 & 0.14 & 0.13 & \textit{0.11}\\
0.25 & 0.35 & 0.33 & 0.35 & 0.32 & 0.36 & 0.17 & 0.17 & 0.17 & 0.18 & 0.17 & \textit{0.16} & 0.21\\
0.50 & 0.53 & 0.51 & 0.54 & 0.48 & 0.54 & 0.27 & 0.29 & 0.28 & 0.30 & \textit{0.26} & \textit{0.26} & 0.35\\
0.75 & 0.74 & 0.69 & 0.75 & 0.62 & 0.73 & 0.52 & 0.56 & 0.53 & 0.60 & \textit{0.48} & \textit{0.48} & 0.50\\
\bottomrule\\
\multicolumn{13}{c}{Second-order GPS model}\\
\toprule
& \multicolumn{12}{c}{$P=5$}\\
\cline{2-13}
$b$ & VM & VM2 & VMnc & VMnr & VMF & KM & KMnc & FM & FMnc & LGPSM & LGPSMnc & COVnc\\
\midrule
0.00 & 0.17 & 0.15 & 0.18 & 0.14 & 0.19 & 0.14 & 0.15 & 0.14 & 0.15 & 0.13 & 0.14 & \textit{0.04}\\
0.25 & 0.37 & 0.35 & 0.37 & 0.35 & 0.38 & 0.13 & 0.15 & 0.14 & 0.15 & 0.11 & 0.13 & \textit{0.07}\\
0.50 & 0.55 & 0.53 & 0.55 & 0.54 & 0.60 & 0.15 & 0.16 & 0.15 & 0.16 & \textit{0.12} & 0.15 & \textit{0.12}\\
0.75 & 0.70 & 0.69 & 0.70 & 0.70 & 0.76 & 0.17 & 0.20 & 0.18 & 0.19 & \textit{0.15} & 0.20 & 0.17\\
1.00 & 0.83 & 0.82 & 0.83 & 0.86 & 0.90 & 0.22 & 0.26 & 0.23 & 0.26 & \textit{0.19} & 0.27 & 0.24\\ 
& \multicolumn{12}{c}{$P=10$}\\
\cline{2-13}
$b$ & VM & VM2 & VMnc & VMnr & VMF & KM & KMnc & FM & FMnc & LGPSM & LGPSMnc & COVnc\\
\midrule
0.00 & 0.21 & 0.18 & 0.21 & 0.17 & 0.22 & 0.18 & 0.18 & 0.18 & 0.18 & 0.17 & 0.17 & \textit{0.08}\\
0.25 & 0.38 & 0.36 & 0.39 & 0.35 & 0.40 & 0.19 & 0.20 & 0.20 & 0.20 & 0.18 & 0.18 & \textit{0.15}\\ 
0.50 & 0.57 & 0.54 & 0.57 & 0.56 & 0.62 & 0.24 & 0.25 & 0.24 & 0.25 & 0.24 & \textit{0.23} & 0.25\\
0.75 & 0.71 & 0.68 & 0.71 & 0.68 & 0.78 & 0.34 & 0.37 & 0.35 & 0.38 & \textit{0.33} & \textit{0.33} & 0.36\\
1.00 & 0.85 & 0.81 & 0.86 & 0.80 & 0.93 & 0.51 & 0.57 & 0.53 & 0.59 & \textit{0.48} & 0.50 & 0.49\\ 
& \multicolumn{12}{c}{$P=20$}\\
\cline{2-13}
$b$ & VM & VM2 & VMnc & VMnr & VMF & KM & KMnc & FM & FMnc & LGPSM & LGPSMnc & COVnc\\
\midrule
0.00 & 0.21 & 0.19 & 0.22 & 0.18 & 0.23 & 0.22 & 0.22 & 0.22 & 0.22 & 0.20 & 0.20 & \textit{0.11}\\
0.25 & 0.39 & 0.37 & 0.39 & 0.36 & 0.41 & 0.24 & 0.25 & 0.25 & 0.26 & 0.24 & 0.23 & \textit{0.21}\\
0.50 & 0.56 & 0.53 & 0.57 & 0.52 & 0.62 & 0.35 & 0.38 & 0.37 & 0.40 & 0.35 & \textit{0.34} & 0.35\\
0.75 & 0.77 & 0.73 & 0.79 & 0.64 & 0.84 & 0.65 & 0.67 & 0.66 & 0.71 & 0.57 & 0.56 & \textit{0.51}\\
\bottomrule
\end{tabular}
\label{bP5table}
\end{table}

\begin{figure}[!t]
\centering
\includegraphics[scale=0.5]{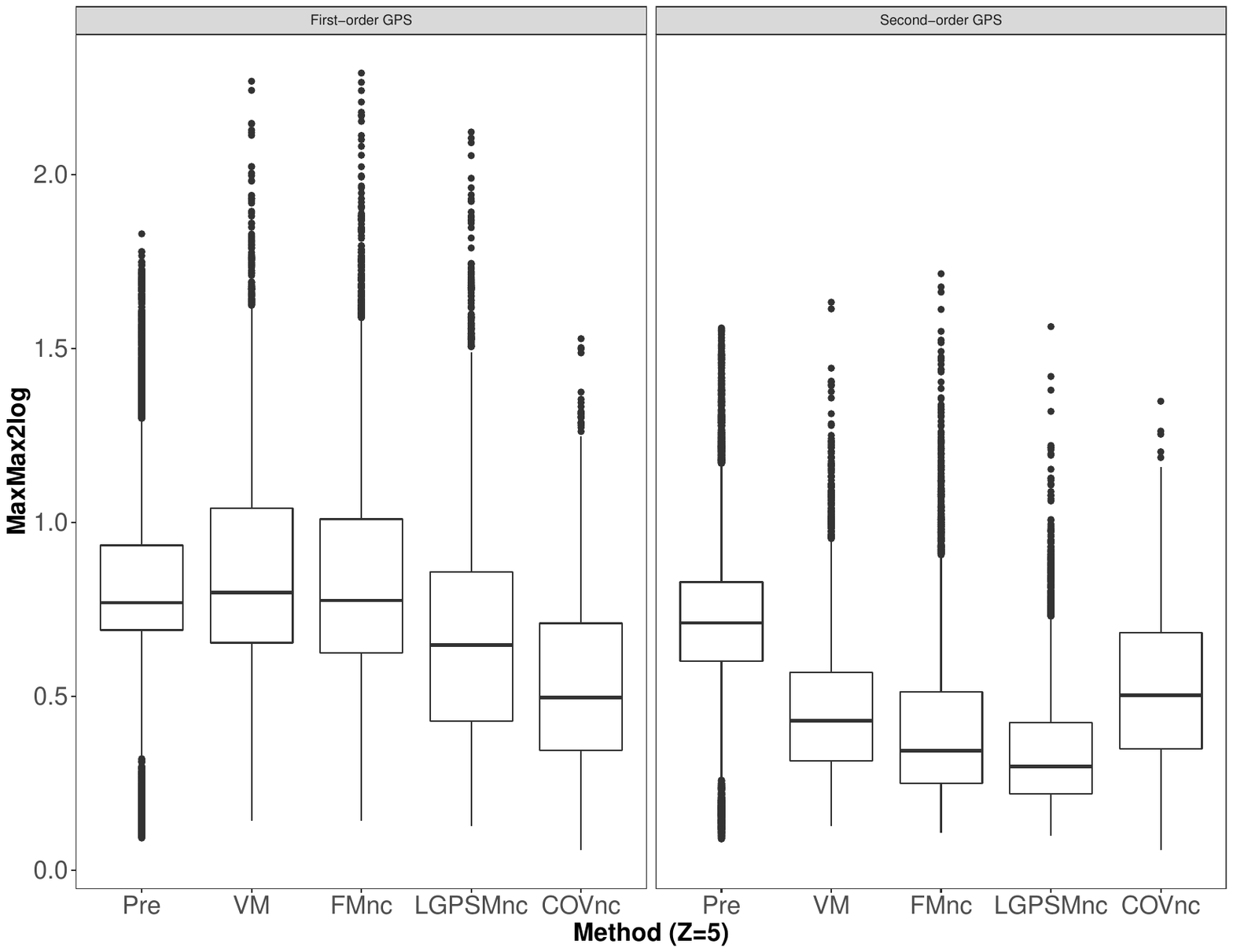}
\caption{$MaxMax2log$ for pre-matched cohort across matching algorithms, $Z=5$}
\label{balance5.log}
\end{figure}

The $MaxMax2log$ for sets generated using each of VM, FMnc, LGPSMnc, and COVnc are depicted in Figure~\ref{balance5.log}. Including the square of the covariates in the GPS model results in smaller $MaxMax2log$ for all algorithms. VM has the highest median $MaxMax2log$ overall for both GPS models. LGPSMnc with second-order GPS model has the lowest median $MaxMax2log$, and 95\% of the configurations with less than 0.69.

Table~\ref{s2s3table5} in the Appendix shows $MaxMax2log$ based on different levels of $\sigma_{2}^{2}$, $\sigma_{3}^{2}$, and GPS order. Similar to $Z=3$, $MaxMax2log$ increases as the absolute difference between $\sigma_{2}^{2}$ and $\sigma_{3}^{2}$ increases. All of the methods that match on the GPS have medians $MaxMax2log$ that are smaller than 0.53 when using the second-order GPS model. COVnc has median $MaxMax2log$ that reaches 0.64.

\subsection{Matching for $Z=10$ treatments}

We implemented VM, FMnc, LGPSMnc, and COVnc for $Z=10$. We only included the non-caliper versions of the latter three algorithms because for $Z=5$ we observed larger biases when these algorithms were implemented with calipers. 

\begin{figure}[!t]
\centering
\includegraphics[scale=0.5]{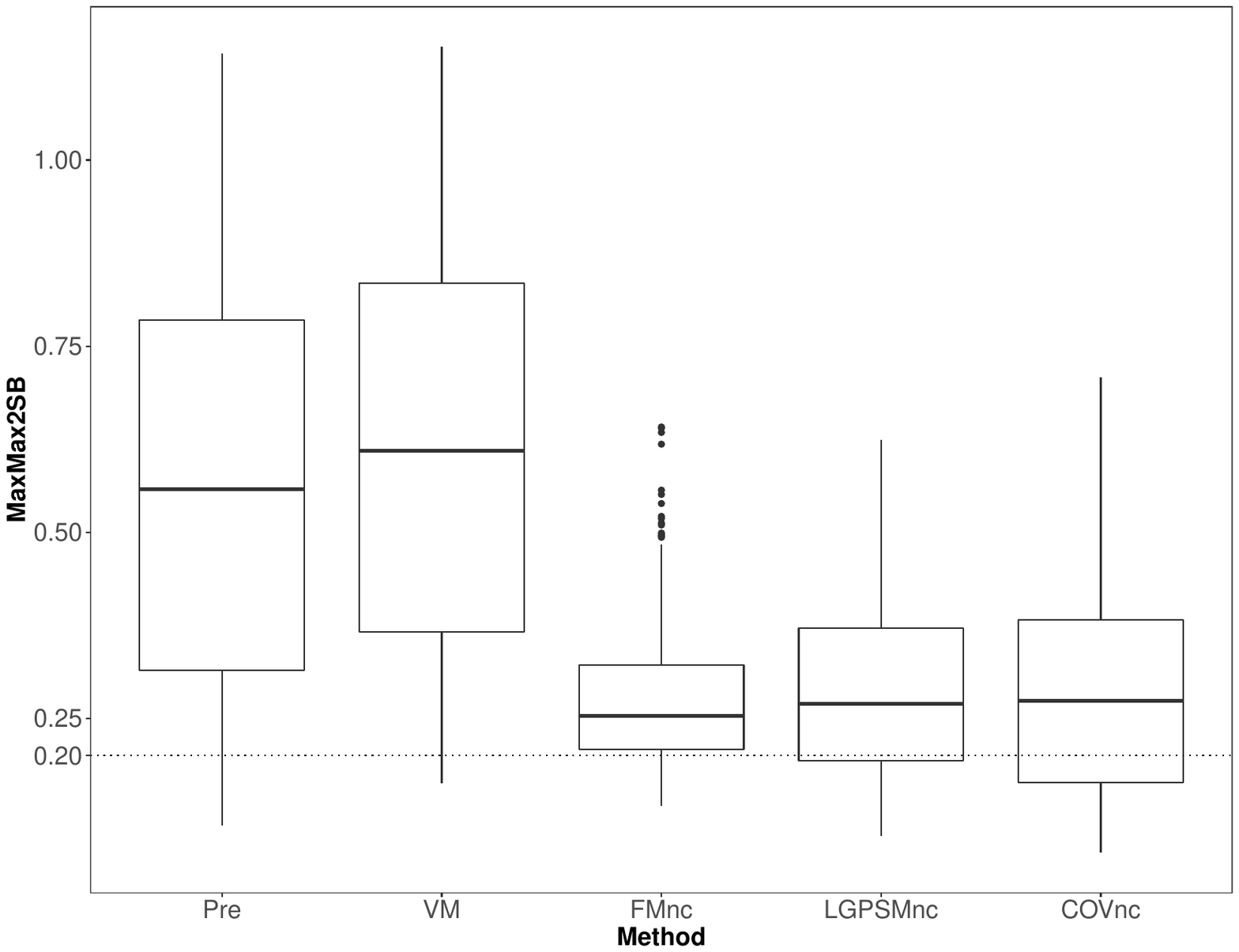}
\caption{$MaxMax2SB$ for pre-matched cohort and by matching algorithms, $Z=10$}
\label{max10}
\end{figure}

The $MaxMax2SB$ for sets generated using each of the four algorithms with a first-order GPS model are depicted in Figure~\ref{max10}. The median $MaxMax2SB$ for VM is significantly larger than the 0.20 cutoff with only 4\% of configurations yielding $MaxMax2SB$ lower than 0.20. The median $MaxMax2SB$ is 0.25 and 0.27 for FMnc and LGPSMnc, respectively. Comparing to configurations with $Z=3$ and $Z=5$, $MaxMax2SB$ is trending upward for LGPSMnc. These results are similar to the results for matching on the Mahalanobis distance of the covariates for binary treatments, where the performance of matching on the Mahalanobis distance deteriorate with increasing $P$ \cite{gu-93, stuart-10}. For FMnc, 19\% of configurations yields $MaxMax2SB$ below 0.20, with an interquartile range of (0.21, 0.32). VM is the only algorithm that used a caliper, and it yields a median $Prop.Matched$ of 0.99. Similar to $Z=3$ and $Z=5$, $MaxMax2SB$ was higher overall when using a second-order GPS model (data not shown). 

Table~\ref{bP10table} shows the median $MaxMax2SB$ for different levels of $b$, $P$, and the two GPS models. Similar to $Z=5$, VM is more sensitive to an increase in $b$. For each algorithm, using either a first- or a second-order GPS model does not appear to influence $MaxMax2SB$ as much as it did for $Z\in\{3, 5\}$. FMnc and GPSnc have medians $MaxMax2SB\leq0.20$ only for $b=0.00$ and $P=10$, though the median $MaxMax2SB$ exceeds 0.30 only for $b\geq0.75$. 

\begin{table}[!t]
\footnotesize
\centering
\caption{Median $MaxMax2SB$ across levels of $b$, $P$, and GPS model order, $Z=10$ (smallest $MaxMax2SB$ in italics)}
\begin{tabular}{c c c c c}
\multicolumn{5}{c}{First-order GPS model}\\
\toprule
& \multicolumn{4}{c}{$P=10$}\\
\cline{2-5}
$b$ & VM & FMnc & LGPSMnc & COVnc\\
\midrule
0.00 & 0.22 & 0.19 & 0.14 & \textit{0.10}\\
0.25 & 0.39 & 0.20 & 0.16 & \textit{0.14}\\ 
0.50 & 0.62 & 0.23 & 0.23 & \textit{0.22}\\
0.75 & 0.84 & \textit{0.29} & 0.31 & 0.31\\
1.00 & 1.03 & \textit{0.37} & 0.42 & 0.42\\ 
& \multicolumn{4}{c}{$P=20$}\\
\cline{2-5}
$b$ & VM & FMnc & LGPSMnc & COVnc\\
\midrule
0.00 & 0.25 & 0.21 & 0.21 & \textit{0.19}\\
0.25 & 0.43 & \textit{0.24} & \textit{0.24} & 0.26\\
0.50 & 0.65 & \textit{0.30} & 0.31 & 0.38\\
0.75 & 0.87 & \textit{0.42} & 0.44 & 0.53\\
\bottomrule\\
\multicolumn{5}{c}{Second-order GPS model}\\
\toprule
& \multicolumn{4}{c}{$P=10$}\\
\cline{2-5}
$b$ & VM & FMnc & LGPSMnc & COVnc\\
\midrule
0.00 & 0.23 & 0.20 & 0.18 & \textit{0.09}\\
0.25 & 0.37 & 0.21 & 0.21 & \textit{0.14}\\ 
0.50 & 0.60 & 0.24 & 0.29 & \textit{0.22}\\
0.75 & 0.80 & \textit{0.28} & 0.37 & 0.31\\
1.00 & 0.99 & \textit{0.36} & 0.47 & 0.42\\ 
& \multicolumn{4}{c}{$P=20$}\\
\cline{2-5}
$b$ & VM & FMnc & LGPSMnc & COVnc\\
\midrule
0.00 & 0.29 & 0.26 & 0.24 & \textit{0.18}\\
0.25 & 0.43 & 0.28 & 0.28 & \textit{0.25}\\
0.50 & 0.65 & \textit{0.33} & 0.36 & 0.38\\
0.75 & 0.87 & \textit{0.48} & \textit{0.48} & 0.52\\
\bottomrule
\end{tabular}
\label{bP10table}
\end{table}

\begin{figure}[!t]
\centering
\includegraphics[scale=0.5]{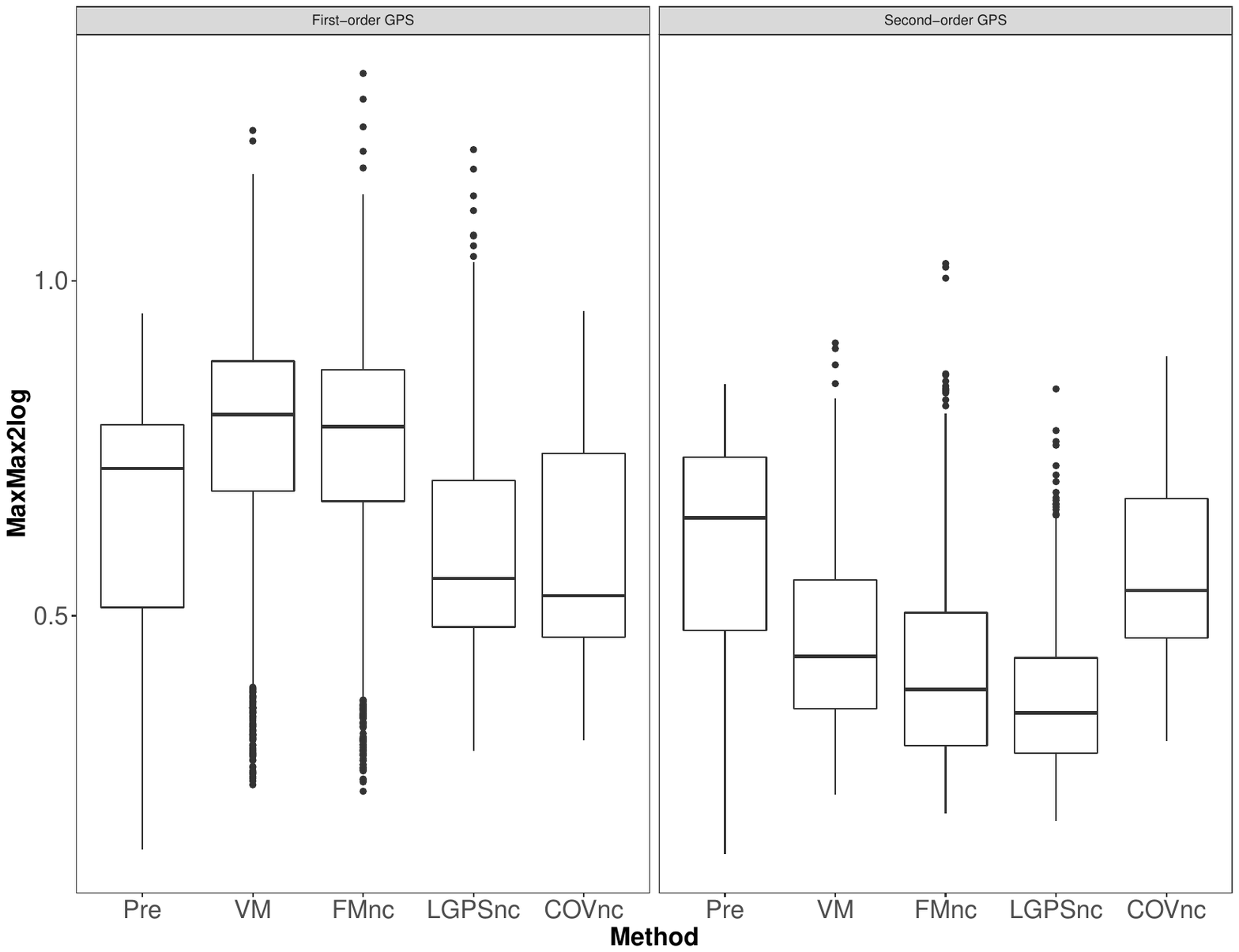}
\caption{$MaxMax2log$ for pre-matched cohort across matching algorithms, $Z=10$}
\label{balance10.log}
\end{figure}

The $MaxMax2log$ for sets generated using each of VM, FMnc, LGPSMnc, and COVnc are depicted in Figure~\ref{balance10.log}. Results are similar to those observed for $Z=5$, where the medians $MaxMax2log$ of VM, FMnc, and LGPSMnc are smaller when using a second-order GPS model, and the $MaxMax2log$ for LGPSMnc is the smallest.

\section{An application to the evaluation of nursing home performance}

\subsection{Nursing home data}

The evaluation of providers' performance based on patient outcomes has played an important role in the analysis and development of health care programs in the United States \cite{ash-12}. To estimate the effects of providers on patients' outcomes, the populations of patients should be similar across all providers, also referred to as similar ``case-mix." While randomly assigning patients to different providers would be ideal, it would not be practical because patients may seek a provider that specializes in a particular condition \cite{huang-05}. 

Rehospitalization rate is a common measure to compare the performance and quality of skilled nursing facilities (SNF) \cite{ouslander-10, ouslander-11}. One possible method to compare SNF rehospitilization rate is risk adjustment using regression modeling \cite{huang-05, neuman-14}. However, regression-based risk adjustment is limited to cases for which the regression model is correctly specified or the covariates are balanced across the different providers \cite{rubin-79, dehejia-99, gutman-13}. Because the risk model is generally unknown and may suffer from model misspecification, it is important to ensure that the distribution of patients across providers is similar in order to avoid extrapolation and biased estimates \cite{ash-12}. 

A different method to compare the rehospitalization rate among providers is to estimate each patient's outcome if they attended each of the available SNFs. To do this, one could match patients admitted to different SNFs who have similar sets of covariates, and impute patients' missing rehospitalization status for an unadmitted SNF using the observed rehospitalization values of the matched units. 

We apply the current and newly proposed matching algorithms to a dataset consisting of Medicare enrollees linked to the Minimum Data Set (MDS) who were discharged from Rhode Island Hospital (RIH) over a nine-year period beginning on January 2, 1999, and assigned to a nearby nursing home in Rhode Island or Massachusetts. The outcome of interest is whether a patient was re-admitted to RIH within 30 days of their initial discharge. For each patient, demographic and clinical characteristics at discharge were recorded. The demographic characteristics include patient's age, gender, and year of admission. The clinical characteristics include the primary ICD-9 diagnosis code for the hospital admission. We have also included in the analysis inflation-adjusted reimbursement for the hospital stay which serves as a proxy for the intensiveness of the stay, as well as intensive care unit (ICU) use and SNF use in the 120 days prior to hospitalization. More than half of the 257 nursing homes in the dataset had fewer than ten patients. We focus on the five nursing homes with the most patients. The number of patients in these SNFs ranges from 670 to 1211, and we define the reference group as the nursing home with the largest number of patients (referred to as ``nursing home 1").

\begin{figure}[!t]
\centering
\includegraphics[scale=0.6]{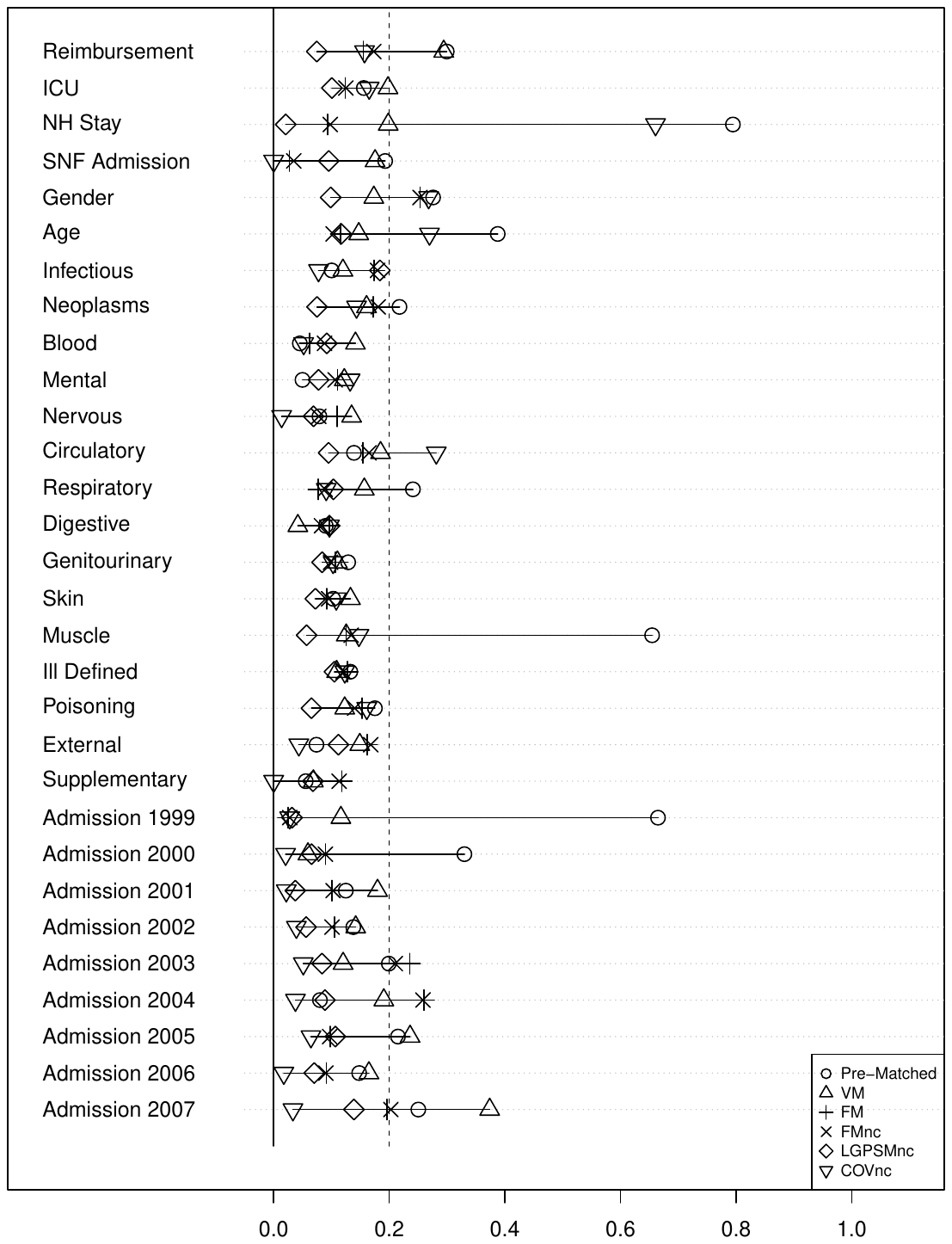}
\caption{$Max2SB_{p}$ for baseline covariates across matching algorithms ($Z=5$)}
\label{balance.max}
\end{figure}

We estimated the GPS vector using a multinomial logistic regression model and generated matched sets using five of the algorithms examined in previous sections: VM, FM, FMnc, LGPSMnc, and COVnc. Figure~\ref{balance.max} reports the maximum absolute standardized pairwise bias, $Max2SB_{p}$, for each of the 30 baseline covariates in the original, unmatched sample, and for the five matched samples generated by the aforementioned procedures. In the original sample, 12 of the 30 covariates have $Max2SB_{p}$ that exceed 0.20, with the largest being prior nursing home use (0.77). LGPSMnc yields the most improved balance with 29 out of 30 covariates having $Max2SB_{p}$ smaller than 0.20 in the matched cohort, while FM and FMnc yield 27 out of 30 covariates having $Max2SB_{p}$ smaller than 0.20. 

Table~\ref{nh.overall.max} shows $MaxMax2SB$ and $Prop.Matched$ across the five algorithms, compared to the pre-matched cohort. LGPSMnc yields the lowest $MaxMax2SB$ of 0.23, followed by FMnc and FM with 0.24 and 0.25, respectively. VM and FM, the two algorithms which used a caliper, retain close to 100\% of the reference group units in the matched cohort. Matching on the Mahalanobis distance of the covariates, COVnc, yields the largest $MaxMax2SB$. 

\begin{table}[!htbp]
\centering
\caption{$MaxMax2SB$ and $Prop.Matched$ for pre-matched and matched samples, nursing home data}
\begin{tabular}{l c c c}
\toprule
Method && $MaxMax2SB$ & $Prop.Matched$\\
\midrule
Pre-matched && 0.77 & --\\
VM && 0.30 & 0.99\\
FM && 0.25 & 0.96\\
FMnc && 0.24 & 1.00\\
LGPSMnc && 0.23 & 1.00\\
COVnc && 0.61 & 1.00\\
\bottomrule
\end{tabular}
\label{nh.overall.max}
\end{table}

Estimated differences in 30-day patient rehospitalization rates are given in Table~\ref{outcomes5} using each of the five matching algorithms for the five nursing homes. Pairwise comparisons between nursing homes 1 and 4 yielded uniformly positive and statistically significant differences in 30-day rehospitalization rates, with the effect size ranging from a 4.42\% (VM) to a 7.28\% (COVnc) higher rehospitalization rate for nursing home 1. Pairwise comparisons between nursing homes 1 and 2 yielded uniformly small and non-statistically significant differences in 30-day rehospitalization rates. 

\begin{table}[!htbp]
\centering
\caption{Estimated outcomes (standard errors in parenthesis) for nursing home data, $Z=5$}
\begin{tabular}{l c c c c}
\toprule
& \multicolumn{4}{c}{Nursing Home Comparison}\\
\cline{2-5}\\
Algorithm & 1 vs 2 & 1 vs 3 & 1 vs 4 & 1 vs 5\\
\midrule
VM & 0.02 $(0.02)$ & -0.02 $(0.02)$ & 0.04 $(0.017)^{*}$ & 0.04 $(0.02)^{*}$\\
FM & 0.00 $(0.02)$ & 0.01 $(0.02)$ & 0.06 $(0.01)^{*}$ & 0.04 $(0.01)^{*}$\\
FMnc & 0.00 $(0.02)$ & 0.01 $(0.02)$ & 0.06 $(0.01)^{*}$ & 0.04 $(0.01)^{*}$\\
LGPSMnc & 0.01 $(0.02)$ & 0.06 $(0.02)^{*}$ & 0.07 $(0.02)^{*}$ & 0.02 $(0.02)$\\
COVnc & 0.03 $(0.02)$ & 0.04 $(0.02)^{*}$ & 0.07 $(0.02)^{*}$ & 0.06 $(0.02)^{*}$\\
\midrule
\multicolumn{2}{c}{$^{*}$Significant at the 0.05 level.} & & &
\end{tabular}
\label{outcomes5}
\end{table}

Notable differences in estimated outcomes obtained from different matching algorithms appear when comparing nursing homes 1 and 3, and nursing homes 1 and 5. When comparing nursing homes 1 and 3, FM and FMnc yield smaller pairwise differences in 30-day rehospitalization rates in comparison to the other methods. When comparing nursing homes 1 and 5, LGPSMnc yields a smaller pairwise difference in 30-day rehospitalization rates, while VM, FM, and FMnc all yield an estimated difference of approximately 3.7\%. 

We further detail the comparisons of nursing home 1 versus nursing homes 2 through 5 by comparing standardized pairwise bias for each covariate. Tables~\ref{sb12}--\ref{sb15} in the Appendix compare $SB_{p1k}$, $k\in\{2,3,4,5\}$, for each matching algorithm. In matching nursing home 1 patients to nursing home 3 patients, FM and FMnc yield greater $SB_{p13}$ for the reimbursement and admission years 2003 and 2007, and COVnc yields greater $SB_{p13}$ for gender, age, and the circulatory disease and poisoning indicators. In matching nursing home 1 patients to nursing home 5 patients, FM and FMnc yield greater $SB_{p15}$ for the ICU and gender covariates, and COVnc yields greater $SB_{p15}$ for the age, ICU, circulatory disease indicator, and poisoning indicator, in comparison to the other methods. LGPSMnc yields similar or smaller $SB_{p13}$ and $SB_{p15}$ for all covariates, findings that are similar to the simulation results for $Z=5$, in which LGPSMnc displays smaller $MaxMax2SB$ on average. With the exception of COVnc, $SB_{p12}$ and $SB_{p14}$ are less than 0.20 for all of the covariates and for all of the matching algorithms that were examined.

\section{Discussion}

Many applications in public health, medicine, and social sciences involve comparing multiple treatment groups. Matching is a useful tool for researchers looking to make causal statements between treatment groups. We proposed several new matching algorithms when evaluating multiple treatments. We compared the performance of the newly proposed algorithms to previously proposed matching algorithms in terms of bias reduction in the covariates' distributions in different treatment groups. 

Vector matching (VM) is shown to result in the lowest covariate bias in comparison to other methods when estimating causal effects with three treatments, similar to the findings for three treatments in Lopez and Gutman (2017) \cite{lopez-15}. Using simulations, we examine several extensions of VM that use different clustering methods, distance measures, and calipers. When comparing three treatments, VM without replacement has the smallest bias; however, it also discards a large number of units receiving the reference treatment, thus sacrificing generalizability for minor gains in bias reductions. Moreover, as the number of treatment groups increases, VM and its extensions based on the linear GPS have the worst performance in terms of bias reduction compared to the other methods that were examined. 

KM and FM are matching algorithms that match on the Mahalanobis distance of pairs of GPS vector components, within strata created by $k$-means or fuzzy clustering of the remaining components, respectively. While simulations show that FM had slightly better performance in terms of balancing covariates, the differences in performance were slight, especially when no caliper was used (KMnc and FMnc). The proportion of reference group units matched is generally higher when using FM, because fuzzy clustering allows for units to be assigned to multiple clusters. Each of these algorithms perform significantly better than VM-like algorithms for five or more treatments, indicating that matching on the Mahalanobis distance within clusters results in better covariate balance than using the linear GPS. 

For three and five treatment groups, when there is a small number of covariates or the bias of the covariates is small, matching on the Mahalanobis distance of the covariates (COVnc) provides the best method in terms of bias reduction. However, the performance of this method deteriorates significantly as the number of covariates with initial biases increases. Thus, this method is not recommended for many practical applications, where a large number of covariates with different degree of initial biases is expected. Matching on the Mahalanobis distance of the logit GPS vector with and without a caliper (LGPSM and LGPSMnc) generally have the largest reduction in bias while maintaining most of the units in the reference group. Matching with a caliper on the Mahalanobis distance of the logit GPS vector (LGPSM) has slightly lower average bias, but it discards more units than matching without a caliper. FM and FMnc have similar trends to LGPSM and LGPSMnc, but with slightly larger covariates' bias. 

For 10 treatment groups, FMnc displays the largest reduction in covariates' bias. This implies that as the number of treatments increases, matching on the Mahalanobis distance of the logit GPS vector becomes less effective. Thus, our recommendation with more than 5 treatments is to rely on methods such as FM or FMnc that partition the cohort into clusters using fuzzy clustering and match units using the Mahalanobis distance within these clusters. In general, as the number of treatments increases, balancing the covariates across all treatment groups is a harder task and imbalances across treatment groups may still remain even after matching. A possible limitation with matching for 10 or more treatment groups is that because of the large number of comparisons being made, it can be difficult to detect significant differences in treatment outcomes after adjustments for multiple comparisons. This problem arises for both randomized and non-randomized studies and it depends on the overall effect sizes, variability of the outcomes in the populations, and the number of units within the different treatment groups.

When matching on a GPS model that includes squared covariates, larger biases in the original covariates are observed, in comparison to a GPS model that only included the covariates. This is because the GPS model with squared terms balances more variables. However, when using a GPS model that includes squared terms, the variance ratios between different treatment groups are significantly lower. For binary treatment, differences in covariates' variance are shown to have large influence on the operating characteristics of causal estimates \cite{gutman-13}. Thus, it is important to ensure that the ratios of variances are similar across treatment groups. Our simulations show that this can be accomplished by including squared terms in the GPS model with relatively minor increases in covariates' biases. 

Applying the different matching algorithms to examine the effect of admission to one of five SNFs reinforced the simulation results. Matching on the Mahalanobis distance of the logit GPS vector without a caliper has the largest reduction in bias while maintaining all of the units in the largest SNF. Based on these results we can identify that residing in certain SNFs can result in a lower chance of being readmitted to a hospital within 30 days. One limitation of this analysis is that it assumes that all of the covariates that influence admission to one of the SNFs are available, or formally, that the assignment mechanism is unconfounded. This is a limitation of any observational study, and it will generally have a larger effect on many of the currently employed risk adjustment techniques as well. Developing sensitivity analyses to this assumption with multiple treatment groups is an area of future research. In addition, health provider profiling generally involves the comparison of thousands of SNFs simultaneously, and development of matching methods in this scenario is also an area of further research. Nonetheless, most patients can only choose from a limited number of possible SNFs. Thus, the proposed matching procedures can provide a viable analysis option for estimating the causal effects of residing in one SNF versus others.

In conclusion, we compared several new and previously proposed matching procedures when attempting to estimate the causal effects from observational studies with multiple treatments. Based on our simulations, for 3-5 treatments matching on the Mahalanobis distance of the logit GPS vector with (LGPSM) or without caliper (LGPSMnc) provide the largest reduction in initial covariate bias. However, as the number of treatments increases, matching methods that are based on the Mahalanobis distance and fuzzy clustering (FM and FMnc) provide better reduction in initial covariates' bias. In addition, matching methods that include the squared term of the covariates in the GPS model result in greater reduction in covariates' variance ratios between treatment groups at the expense of slightly larger covariates' bias.

\subsection*{Acknowledgements} 

This research was supported through a Patient-Centered Outcomes Research Institute (PCORI) Award ME-1403-12104. Disclaimer: All statements in this report, including its findings and conclusions, are solely those of the authors and do not necessarily represent the views of the Patient-Centered Outcomes Research Institute (PCORI), its Board of Governors or Methodology Committee. 

\subsection*{Financial disclosure}

None reported.

\subsection*{Conflict of interest}

The authors declare no potential conflict of interests.

\subsection*{Supporting Information}

Data are available in a synthetic format at \texttt{https://github.com/ScotinaStats}, along with sample R code.

\newpage

\bibliographystyle{wileyj}
\bibliography{references}

\newpage

\section*{Appendix}

\section*{Additional Tables and Figures}

\begin{table}[!htbp]
\centering
\caption{Median $MaxMax2log$ across levels of $\sigma_{2}^{2}$, $\sigma_{3}^{2}$, and GPS model order, $Z=3$}
\begin{tabular}{l l l | c c c c}
\toprule
& $\sigma_{2}^{2}$ & $\sigma_{3}^{2}$ & VM & FMnc & LGPSMnc & COVnc\\
\midrule
\multirow{9}{*}{First-order GPS} & 0.5 & 0.5 & 0.77 & 0.76 & 0.77 & 0.67\\
& 0.5 & 1.0 & 0.78 & 0.77 & 0.77 & 0.66\\
& 0.5 & 2.0 & 1.40 & 1.38 & 1.38 & 0.70\\
& 1.0 & 0.5 & 0.77 & 0.77 & 0.76 & 0.63\\
& 1.0 & 1.0 & 0.27 & 0.25 & 0.25 & 0.34\\
& 1.0 & 2.0 & 0.81 & 0.80 & 0.80 & 0.38\\
& 2.0 & 0.5 & 1.39 & 1.36 & 1.36 & 0.66\\
& 2.0 & 1.0 & 0.81 & 0.80 & 0.80 & 0.38\\
& 2.0 & 2.0 & 0.81 & 0.80 & 0.80 & 0.20\\
\midrule
\multirow{9}{*}{Second-order GPS} & 0.5 & 0.5 & 0.37 & 0.38 & 0.37 & 0.61\\
& 0.5 & 1.0 & 0.34 & 0.34 & 0.33 & 0.61\\
& 0.5 & 2.0 & 0.37 & 0.33 & 0.32 & 0.63\\
& 1.0 & 0.5 & 0.33 & 0.32 & 0.32 & 0.59\\
& 1.0 & 1.0 & 0.22 & 0.20 & 0.20 & 0.34\\
& 1.0 & 2.0 & 0.28 & 0.23 & 0.23 & 0.38\\
& 2.0 & 0.5 & 0.35 & 0.31 & 0.30 & 0.61\\
& 2.0 & 1.0 & 0.28 & 0.24 & 0.24 & 0.38\\
& 2.0 & 2.0 & 0.23 & 0.22 & 0.22 & 0.20\\
\bottomrule
\end{tabular}
\label{s2s3table3}
\end{table}

\newpage

\begin{table}[!htbp]
\centering
\caption{Median $MaxMax2log$ across levels of $\sigma_{2}^{2}$, $\sigma_{3}^{2}$, and GPS model order, $Z=5$}
\begin{tabular}{l l l | c c c c}
\toprule
& $\sigma_{2}^{2}$ & $\sigma_{3}^{2}$ & VM & FMnc & LGPSMnc & COVnc\\
\midrule
\multirow{9}{*}{First-order GPS} & 0.5 & 0.5 & 0.75 & 0.69 & 0.63 & 0.65\\
& 0.5 & 1.0 & 0.75 & 0.71 & 0.61 & 0.65\\
& 0.5 & 2.0 & 1.25 & 1.17 & 0.98 & 0.67\\
& 1.0 & 0.5 & 0.76 & 0.72 & 0.59 & 0.61\\
& 1.0 & 1.0 & 0.32 & 0.31 & 0.28 & 0.36\\
& 1.0 & 2.0 & 0.77 & 0.76 & 0.62 & 0.41\\
& 2.0 & 0.5 & 1.25 & 1.20 & 0.95 & 0.64\\
& 2.0 & 1.0 & 0.77 & 0.75 & 0.62 & 0.42\\
& 2.0 & 2.0 & 0.78 & 0.76 & 0.63 & 0.41\\
\midrule
\multirow{9}{*}{Second-order GPS} & 0.5 & 0.5 & 0.42 & 0.40 & 0.34 & 0.60\\
& 0.5 & 1.0 & 0.44 & 0.37 & 0.31 & 0.61\\
& 0.5 & 2.0 & 0.53 & 0.40 & 0.35 & 0.64\\
& 1.0 & 0.5 & 0.40 & 0.36 & 0.31 & 0.58\\
& 1.0 & 1.0 & 0.29 & 0.26 & 0.24 & 0.36\\
& 1.0 & 2.0 & 0.44 & 0.29 & 0.26 & 0.41\\
& 2.0 & 0.5 & 0.53 & 0.39 & 0.34 & 0.63\\
& 2.0 & 1.0 & 0.46 & 0.30 & 0.26 & 0.42\\
& 2.0 & 2.0 & 0.36 & 0.29 & 0.26 & 0.43\\
\bottomrule
\end{tabular}
\label{s2s3table5}
\end{table}

\newpage

\begin{figure}[!htbp]
\centering
\includegraphics[scale=0.5]{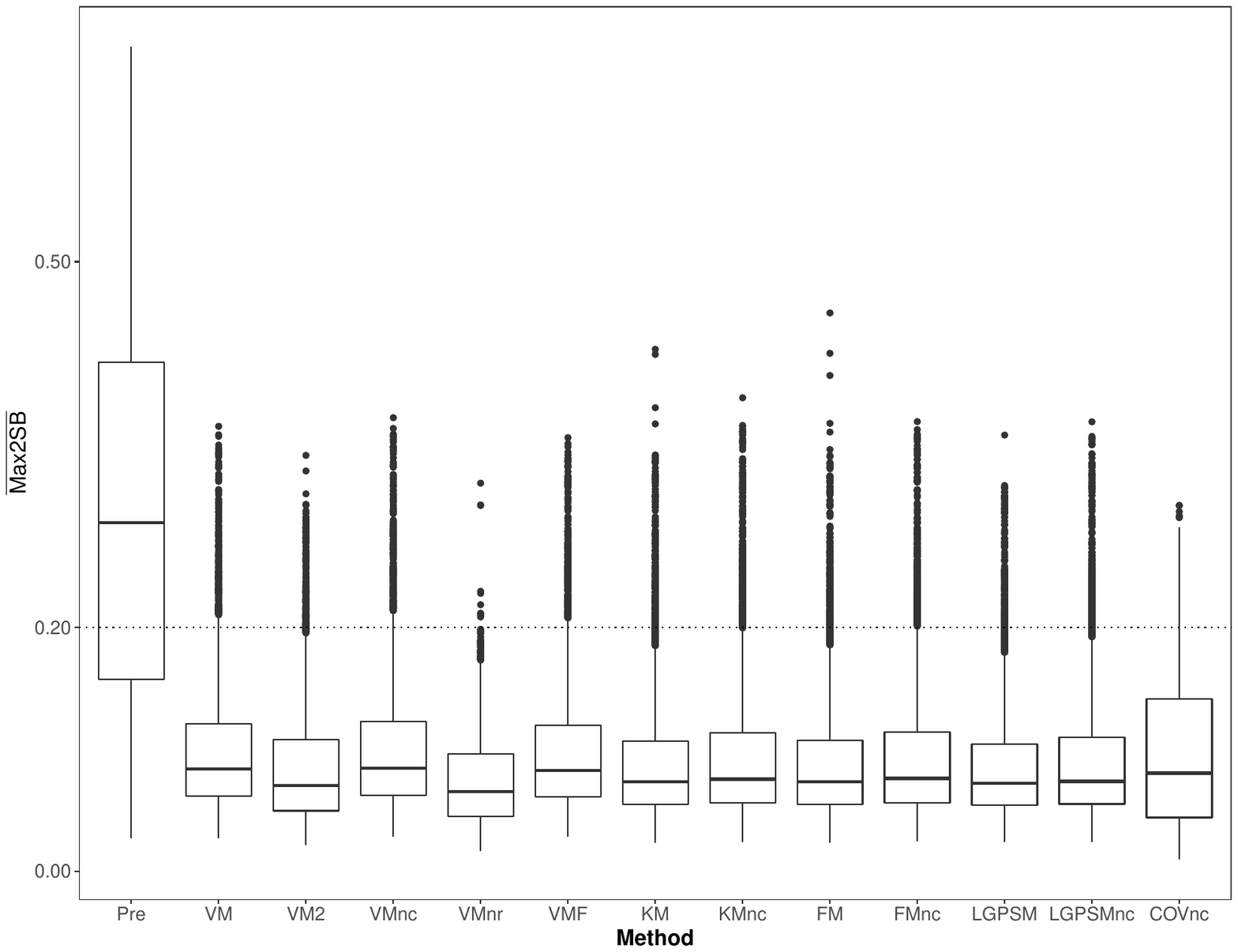}
\caption{$\overline{Max2SB}$ for pre-matched cohort and by matching algorithms, $Z=3$}
\label{mean3}
\end{figure}

\newpage

\begin{figure}[!htbp]
\centering
\includegraphics[scale=0.5]{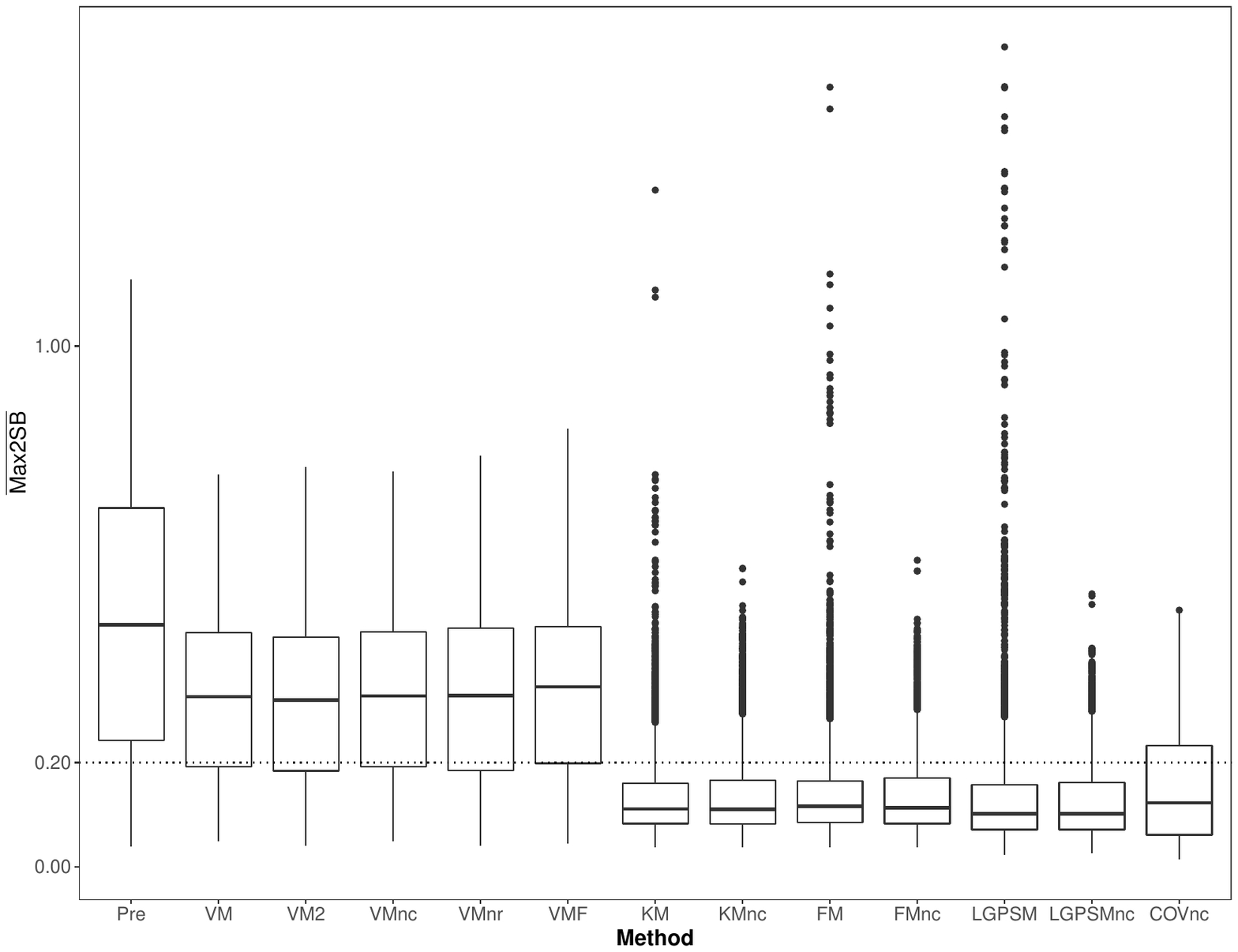}
\caption{$\overline{Max2SB}$ for pre-matched cohort and by matching algorithms, $Z=5$}
\label{mean5}
\end{figure}

\newpage

\begin{figure}[!htbp]
\centering
\includegraphics[scale=0.5]{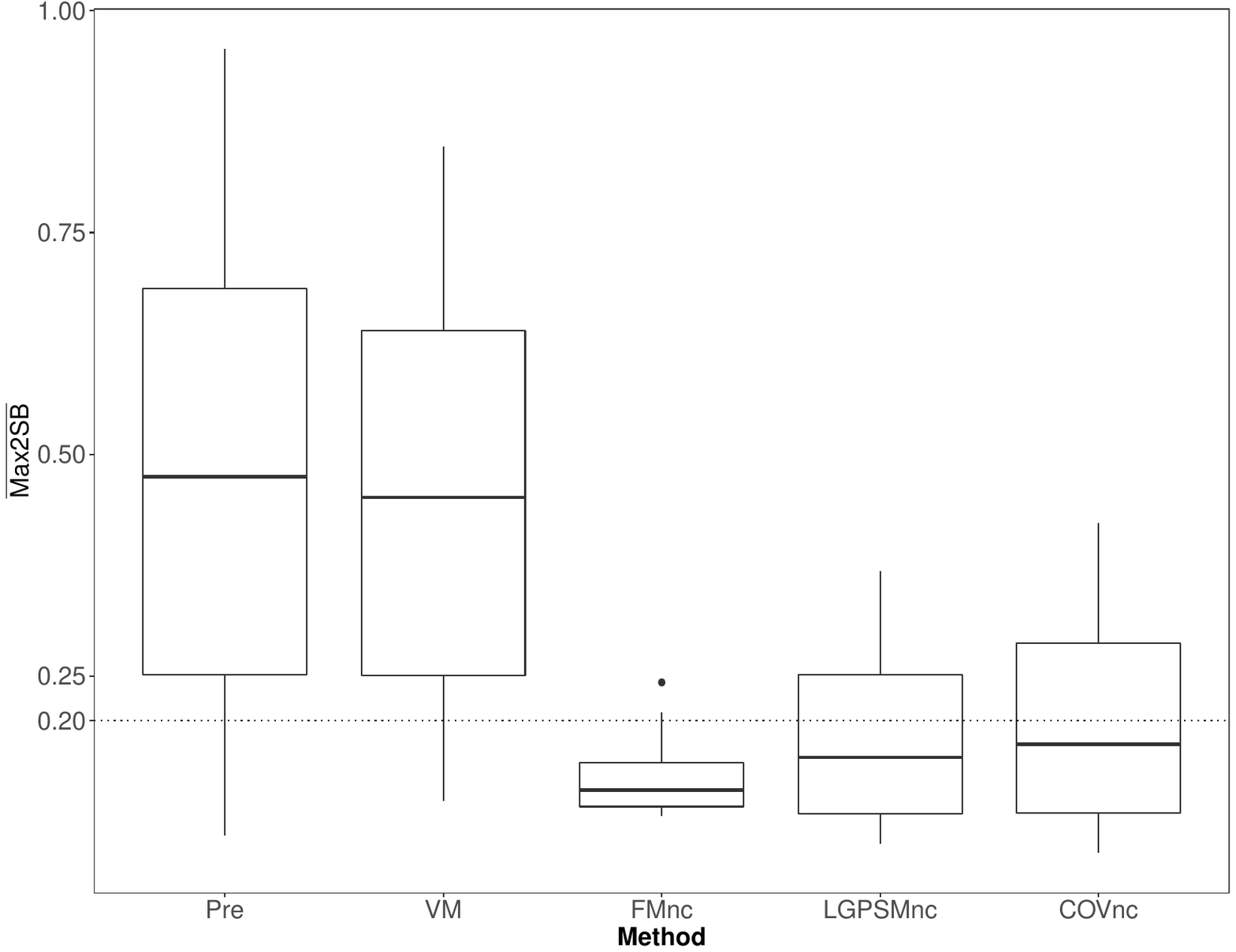}
\caption{$\overline{Max2SB}$ for pre-matched cohort and by matching algorithms, $Z=10$}
\label{mean10}
\end{figure}

\newpage

\begin{table}[!htbp]
\centering
\caption{$SB_{p12}$ for nursing home data}
\begin{tabular}{lrrrrr}
  \toprule
Variable & VM & FM & FMnc & LGPSMnc & COVnc \\ 
  \midrule
Reimbursement & 0.07 & 0.01 & 0.01 & 0.05 & 0.16 \\ 
  ICU & 0.02 & 0.00 & 0.02 & 0.05 & 0.13 \\ 
  NH Stay & 0.02 & 0.01 & 0.01 & 0.03 & 0.07 \\ 
  SNF Admission & 0.10 & 0.03 & 0.03 & 0.00 & 0.00 \\ 
  Gender & 0.05 & 0.01 & 0.01 & 0.02 & 0.23 \\ 
  Age & 0.13 & 0.01 & 0.00 & 0.08 & 0.21 \\ 
  Infectious & 0.00 & 0.05 & 0.06 & 0.03 & 0.00 \\ 
  Neoplasms & 0.00 & 0.07 & 0.07 & 0.01 & 0.02 \\ 
  Blood & 0.06 & 0.02 & 0.03 & 0.06 & 0.02 \\ 
  Mental & 0.02 & 0.01 & 0.02 & 0.00 & 0.09 \\ 
  Nervous & 0.00 & 0.01 & 0.02 & 0.05 & 0.01 \\ 
  Circulatory & 0.03 & 0.02 & 0.01 & 0.04 & 0.20 \\ 
  Respiratory & 0.01 & 0.05 & 0.04 & 0.07 & 0.03 \\ 
  Digestive & 0.00 & 0.00 & 0.01 & 0.07 & 0.04 \\ 
  Genitourinary & 0.02 & 0.01 & 0.02 & 0.02 & 0.03 \\ 
  Skin & 0.05 & 0.06 & 0.06 & 0.03 & 0.05 \\ 
  Muscle & 0.01 & 0.02 & 0.01 & 0.03 & 0.04 \\ 
  Ill Defined & 0.08 & 0.03 & 0.03 & 0.00 & 0.12 \\ 
  Poisoning & 0.02 & 0.01 & 0.00 & 0.01 & 0.09 \\ 
  External & 0.06 & 0.01 & 0.02 & 0.00 & 0.00 \\ 
  Supplementary & 0.03 & 0.02 & 0.04 & 0.05 & 0.00 \\ 
  Admission 1999 & 0.04 & 0.02 & 0.01 & 0.02 & 0.00 \\ 
  Admission 2000 & 0.04 & 0.01 & 0.01 & 0.03 & 0.00 \\ 
  Admission 2001 & 0.02 & 0.05 & 0.05 & 0.03 & 0.01 \\ 
  Admission 2002 & 0.03 & 0.02 & 0.02 & 0.02 & 0.02 \\ 
  Admission 2003 & 0.03 & 0.02 & 0.01 & 0.07 & 0.01 \\ 
  Admission 2004 & 0.01 & 0.03 & 0.03 & 0.02 & 0.02 \\ 
  Admission 2005 & 0.08 & 0.03 & 0.03 & 0.02 & 0.03 \\ 
  Admission 2006 & 0.07 & 0.03 & 0.04 & 0.10 & 0.02 \\ 
  Admission 2007 & 0.03 & 0.00 & 0.01 & 0.00 & 0.01 \\ 
   \bottomrule
\end{tabular}
\label{sb12}
\end{table}

\newpage

\begin{table}[!htbp]
\centering
\caption{$SB_{p13}$ for nursing home data}
\begin{tabular}{lrrrrr}
  \toprule
Variable & VM & FM & FMnc & LGPSMnc & COVnc \\ 
  \midrule
Reimbursement & 0.07 & 0.18 & 0.12 & 0.04 & 0.13 \\ 
  ICU & 0.02 & 0.10 & 0.08 & 0.10 & 0.07 \\ 
  NH Stay & 0.00 & 0.06 & 0.06 & 0.03 & 0.23 \\ 
  SNF Admission & 0.05 & 0.01 & 0.01 & 0.03 & 0.00 \\ 
  Gender & 0.04 & 0.02 & 0.01 & 0.06 & 0.12 \\ 
  Age & 0.01 & 0.02 & 0.02 & 0.08 & 0.28 \\ 
  Infectious & 0.02 & 0.05 & 0.04 & 0.06 & 0.07 \\ 
  Neoplasms & 0.06 & 0.06 & 0.04 & 0.02 & 0.04 \\ 
  Blood & 0.05 & 0.00 & 0.01 & 0.01 & 0.01 \\ 
  Mental & 0.02 & 0.03 & 0.03 & 0.00 & 0.08 \\ 
  Nervous & 0.05 & 0.02 & 0.05 & 0.13 & 0.01 \\ 
  Circulatory & 0.04 & 0.01 & 0.02 & 0.01 & 0.11 \\ 
  Respiratory & 0.07 & 0.02 & 0.01 & 0.09 & 0.06 \\ 
  Digestive & 0.03 & 0.06 & 0.05 & 0.06 & 0.09 \\ 
  Genitourinary & 0.06 & 0.02 & 0.02 & 0.05 & 0.01 \\ 
  Skin & 0.06 & 0.07 & 0.05 & 0.00 & 0.08 \\ 
  Muscle & 0.14 & 0.11 & 0.09 & 0.05 & 0.06 \\ 
  Ill Defined & 0.07 & 0.01 & 0.02 & 0.06 & 0.11 \\ 
  Poisoning & 0.04 & 0.01 & 0.01 & 0.03 & 0.15 \\ 
  External & 0.04 & 0.07 & 0.08 & 0.04 & 0.06 \\ 
  Supplementary & 0.19 & 0.13 & 0.11 & 0.03 & 0.00 \\ 
  Admission 1999 & 0.02 & 0.05 & 0.04 & 0.01 & 0.00 \\ 
  Admission 2000 & 0.01 & 0.02 & 0.02 & 0.02 & 0.00 \\ 
  Admission 2001 & 0.01 & 0.03 & 0.04 & 0.02 & 0.02 \\ 
  Admission 2002 & 0.01 & 0.09 & 0.07 & 0.01 & 0.01 \\ 
  Admission 2003 & 0.11 & 0.05 & 0.04 & 0.16 & 0.00 \\ 
  Admission 2004 & 0.08 & 0.18 & 0.17 & 0.08 & 0.01 \\ 
  Admission 2005 & 0.17 & 0.12 & 0.11 & 0.14 & 0.01 \\ 
  Admission 2006 & 0.09 & 0.07 & 0.05 & 0.06 & 0.01 \\ 
  Admission 2007 & 0.12 & 0.14 & 0.13 & 0.12 & 0.02 \\ 
   \bottomrule
\end{tabular}
\label{sb13}
\end{table}

\newpage

\begin{table}[!htbp]
\centering
\caption{$SB_{p14}$ for nursing home data}
\begin{tabular}{lrrrrr}
  \toprule
Variable & VM & FM & FMnc & LGPSMnc & COVnc \\ 
  \midrule
Reimbursement & 0.16 & 0.09 & 0.03 & 0.01 & 0.00 \\ 
  ICU & 0.14 & 0.00 & 0.03 & 0.04 & 0.16 \\ 
  NH Stay & 0.26 & 0.01 & 0.01 & 0.01 & 0.38 \\ 
  SNF Admission & 0.03 & 0.02 & 0.06 & 0.03 & 0.00 \\ 
  Gender & 0.04 & 0.00 & 0.00 & 0.07 & 0.09 \\ 
  Age & 0.13 & 0.01 & 0.03 & 0.05 & 0.06 \\ 
  Infectious & 0.05 & 0.02 & 0.01 & 0.00 & 0.05 \\ 
  Neoplasms & 0.04 & 0.05 & 0.05 & 0.01 & 0.04 \\ 
  Blood & 0.04 & 0.04 & 0.03 & 0.07 & 0.03 \\ 
  Mental & 0.09 & 0.04 & 0.04 & 0.03 & 0.13 \\ 
  Nervous & 0.01 & 0.03 & 0.03 & 0.00 & 0.00 \\ 
  Circulatory & 0.12 & 0.03 & 0.03 & 0.13 & 0.29 \\ 
  Respiratory & 0.05 & 0.02 & 0.02 & 0.01 & 0.10 \\ 
  Digestive & 0.04 & 0.01 & 0.00 & 0.09 & 0.08 \\ 
  Genitourinary & 0.01 & 0.07 & 0.07 & 0.03 & 0.10 \\ 
  Skin & 0.08 & 0.08 & 0.07 & 0.11 & 0.11 \\ 
  Muscle & 0.00 & 0.06 & 0.03 & 0.03 & 0.08 \\ 
  Ill Defined & 0.03 & 0.08 & 0.08 & 0.04 & 0.12 \\ 
  Poisoning & 0.02 & 0.02 & 0.02 & 0.05 & 0.06 \\ 
  External & 0.03 & 0.03 & 0.02 & 0.02 & 0.04 \\ 
  Supplementary & 0.01 & 0.06 & 0.05 & 0.00 & 0.00 \\ 
  Admission 1999 & 0.06 & 0.02 & 0.02 & 0.06 & 0.01 \\ 
  Admission 2000 & 0.06 & 0.00 & 0.01 & 0.04 & 0.02 \\ 
  Admission 2001 & 0.16 & 0.05 & 0.05 & 0.05 & 0.00 \\ 
  Admission 2002 & 0.01 & 0.03 & 0.02 & 0.05 & 0.00 \\ 
  Admission 2003 & 0.05 & 0.05 & 0.03 & 0.07 & 0.02 \\ 
  Admission 2004 & 0.03 & 0.02 & 0.02 & 0.09 & 0.00 \\ 
  Admission 2005 & 0.10 & 0.12 & 0.11 & 0.06 & 0.00 \\ 
  Admission 2006 & 0.04 & 0.03 & 0.04 & 0.03 & 0.01 \\ 
  Admission 2007 & 0.05 & 0.10 & 0.10 & 0.05 & 0.01 \\  
   \bottomrule
\end{tabular}
\label{sb14}
\end{table}

\newpage

\begin{table}[!htbp]
\centering
\caption{$SB_{p15}$ for nursing home data}
\begin{tabular}{lrrrrr}
  \toprule
Variable & VM & FM & FMnc & LGPSMnc & COVnc \\ 
  \midrule
Reimbursement & 0.07 & 0.01 & 0.07 & 0.04 & 0.12 \\ 
  ICU & 0.07 & 0.10 & 0.12 & 0.05 & 0.16 \\ 
  NH Stay & 0.05 & 0.06 & 0.06 & 0.01 & 0.02 \\ 
  SNF Admission & 0.04 & 0.01 & 0.00 & 0.03 & 0.00 \\ 
  Gender & 0.04 & 0.13 & 0.12 & 0.01 & 0.04 \\ 
  Age & 0.05 & 0.07 & 0.04 & 0.04 & 0.16 \\ 
  Infectious & 0.03 & 0.01 & 0.01 & 0.13 & 0.05 \\ 
  Neoplasms & 0.07 & 0.07 & 0.07 & 0.08 & 0.10 \\ 
  Blood & 0.04 & 0.05 & 0.06 & 0.03 & 0.05 \\ 
  Mental & 0.01 & 0.02 & 0.01 & 0.02 & 0.08 \\ 
  Nervous & 0.01 & 0.10 & 0.10 & 0.06 & 0.02 \\ 
  Circulatory & 0.07 & 0.03 & 0.03 & 0.07 & 0.22 \\ 
  Respiratory & 0.02 & 0.03 & 0.03 & 0.01 & 0.05 \\ 
  Digestive & 0.05 & 0.07 & 0.08 & 0.03 & 0.07 \\ 
  Genitourinary & 0.05 & 0.02 & 0.02 & 0.03 & 0.00 \\ 
  Skin & 0.05 & 0.04 & 0.05 & 0.01 & 0.10 \\ 
  Muscle & 0.02 & 0.02 & 0.04 & 0.01 & 0.07 \\ 
  Ill Defined & 0.02 & 0.09 & 0.08 & 0.05 & 0.09 \\ 
  Poisoning & 0.04 & 0.08 & 0.06 & 0.04 & 0.17 \\ 
  External & 0.00 & 0.02 & 0.03 & 0.02 & 0.01 \\ 
  Supplementary & 0.02 & 0.03 & 0.01 & 0.03 & 0.00 \\ 
  Admission 1999 & 0.03 & 0.01 & 0.01 & 0.02 & 0.01 \\ 
  Admission 2000 & 0.01 & 0.02 & 0.02 & 0.01 & 0.01 \\ 
  Admission 2001 & 0.04 & 0.03 & 0.04 & 0.05 & 0.01 \\ 
  Admission 2002 & 0.05 & 0.06 & 0.05 & 0.02 & 0.02 \\ 
  Admission 2003 & 0.07 & 0.13 & 0.11 & 0.05 & 0.02 \\ 
  Admission 2004 & 0.10 & 0.07 & 0.07 & 0.02 & 0.04 \\ 
  Admission 2005 & 0.14 & 0.13 & 0.13 & 0.00 & 0.03 \\ 
  Admission 2006 & 0.02 & 0.05 & 0.05 & 0.02 & 0.02 \\ 
  Admission 2007 & 0.16 & 0.06 & 0.07 & 0.01 & 0.01 \\ 
   \bottomrule
\end{tabular}
\label{sb15}
\end{table}

\newpage

\section*{Nursing home data codebook}

\begin{itemize}
	\item Reimbursement -- inflation-adjusted reimbursement (2000\$)
	\item ICU -- any ICU during hospitalization
	\item NH Stay -- Nursing home stay within 120 days before inpatient admission
	\item SNF Admission -- SNF admission  (yes/no)
	\item Gender -- beneficiary sex (1=male, 2=female)
	\item Age -- beneficiary age
	\item Infectious -- primary ICD-9 diagnosis 001--139: infectious and parasitic diseases
	\item Neoplasms -- primary ICD-9 diagnosis 140--239: neoplasms
	\item Blood -- primary ICD-9 diagnosis 280--289: diseases of the blood and blood-forming organs
	\item Mental -- primary ICD-9 diagnosis 290--319: mental disorders
	\item Nervous -- primary ICD-9 diagnosis 320--389: diseases of the nervous system and sense organs
	\item Circulatory -- primary ICD-9 diagnosis 390--459: diseases of the circulatory system
	\item Respiratory -- primary ICD-9 diagnosis 460--519: diseases of the respiratory system
	\item Digestive -- primary ICD-9 diagnosis 520--579: diseases of the digestive system
	\item Genitourinary -- primary ICD-9 diagnosis 580--629: diseases of the genitourinary system
	\item Skin -- primary ICD-9 diagnosis 680--709: diseases of the skin and subcutaneous tissue
	\item Muscle -- primary ICD-9 diagnosis 710--739: diseases of the musculoskeletal system and connective tissue
	\item Ill Defined -- primary ICD-9 diagnosis 780--799: symptoms, signs, and ill-defined conditions
	\item Poisoning -- primary ICD-9 diagnosis 800--999: injury and poisoning
	\item External -- primary ICD-9 diagnosis E: external causes of injury
	\item Supplementary -- primary ICD-9 diagnosis V: supplemental classification
	\item Admission 1999 -- Nursing home year of admission in 1999
	\item Admission 2000 -- Nursing home year of admission in 2000
	\item Admission 2001 -- Nursing home year of admission in 2001
	\item Admission 2002 -- Nursing home year of admission in 2002
	\item Admission 2003 -- Nursing home year of admission in 2003
	\item Admission 2004 -- Nursing home year of admission in 2004
	\item Admission 2005 -- Nursing home year of admission in 2005
	\item Admission 2006 -- Nursing home year of admission in 2006
	\item Admission 2007 -- Nursing home year of admission in 2007
\end{itemize}

\end{document}